\documentclass[aps,onecolumn,tightenlines,preprint,11pt,nofootinbib,letterpaper]{revtex4-1}

\usepackage{amsmath}
\usepackage{amssymb}
\usepackage{epsfig}
\usepackage{graphicx}

\usepackage{multirow}
\usepackage{longtable}

\usepackage{bbm}
\usepackage{mathrsfs}

\makeatletter
\renewcommand{\fnum@table}{\textbf{\tablename~\thetable}}
\renewcommand{\fnum@figure}{\textbf{\figurename~\thefigure}}
\makeatother

\newcounter{myenumi}

\renewcommand{\themyenumi}{\roman{myenumi}}
{\end{list}}

\newlength{\myem}
\newcounter{mysubequation}[equation]

\makeatletter
\renewcommand{\section}{\@startsection{section}{1}{0em}{-\baselineskip}%
{\baselineskip}{\normalfont\large\bfseries}}
\renewcommand{\subsection}%
{\@startsection{subsection}{2}{0em}{-0.7\baselineskip}%
{0.7\baselineskip}{\normalfont\bfseries}}
\makeatother

\newcommand{\bi}{\begin{itemize}}
\newcommand{\ei}{\end{itemize}}

\newcommand{\be}{\begin{equation}}
\newcommand{\ee}{\end{equation}}
\newcommand{\bea}{\begin{eqnarray}}
\newcommand{\eea}{\end{eqnarray}}

\newcommand{\ldm}{\Delta m_{31}^2}
\newcommand{\sdm}{\Delta m_{21}^2}
\newcommand{\deltacp}{\delta_{\mathrm{CP}}}

\newcommand{\bra}[1]{\ensuremath{\langle #1 |}}   
\newcommand{\ket}[1]{\ensuremath{| #1 \rangle}}   
\newcommand{\diag}{{\rm diag}} 

\newcommand{\eps}{\varepsilon}
\newcommand{\deltaCP}{\ensuremath{\delta_\mathrm{CP}}}
\newcommand{\dm}[1]{\ensuremath{\Delta m^2_{#1}}}  

\begin{document}

\preprint{VT-IPNAS 10-15}
\preprint{FERMILAB-PUB-10-415-T}

\title{Two experiments for the price of one? \\
The role of the second oscillation maximum in long baseline neutrino experiments}

\author{Patrick Huber}
\email{pahuber@vt.edu}

\affiliation{Department of Physics, Virginia Tech, Blacksburg, VA 24061, USA}

\author{Joachim Kopp}
\email{jkopp@fnal.gov}

\affiliation{Fermilab, Theoretical Physics Department, PO Box 500,
  Batavia, IL 60510, USA}

\date{\today}

\begin{abstract}
  We investigate the quantitative impact that data from the second
  oscillation maximum has on the performance of wide band beam
  neutrino oscillation experiments. We present results for the physics
  sensitivities to standard three flavor oscillation, as well as
  results for the sensitivity to non-standard interactions. The
  quantitative study is performed using an experimental setup similar
  to the Fermilab to DUSEL Long Baseline Neutrino Experiment (LBNE).
  We find that, with the single exception of sensitivity to the mass
  hierarchy, the second maximum plays only a marginal role due to the
  experimental difficulties to obtain a statistically significant and
  sufficiently background-free event sample at low energies. This
  conclusion is valid for both water \v{C}erenkov and liquid argon
  detectors. Moreover, we confirm that non-standard neutrino
  interactions are very hard to distinguish experimentally from
  standard three-flavor effects and can lead to a considerable loss of
  sensitivity to $\theta_{13}$, the mass hierarchy and CP violation.
\end{abstract}
\maketitle
\section{Introduction}
\label{sec:intro}

Neutrino physics has seen a spectacular transition from a collection
of anomalies to a field of precision study with a firmly established
theoretical underpinning. All neutrino flavor transition data, with
the exception of
LSND~\cite{Aguilar:2001ty,*Athanassopoulos:1997er,*Athanassopoulos:1995iw,*Athanassopoulos:1997er}
and
MiniBooNE~\cite{AguilarArevalo:2007it,*AguilarArevalo:2008rc,*AguilarArevalo:2010wv},
can be described by oscillation of three active neutrino flavors, see
{\it e.g.} the reviews in \cite{GonzalezGarcia:2007ib,Maltoni:2004ei}.
Throughout this paper we assume that LSND and MiniBooNE have
explanations which do not affect our results, {\it i.e.}\ they are not
due to neutrino oscillation.

Within the three flavor oscillation framework, experiments have
determined the values of the mass squared differences and the
associated large mixing angles with a precision at the level of a few
percent~\cite{Maltoni:2004ei}. Currently unknown are the size of
$\theta_{13}$, the value of the CP phase $\deltacp$ and the sign of
the atmospheric mass splitting $\dm{31}$, as well whether
$\theta_{23}$ is exactly $\pi/4$, and if not, whether it is larger or
smaller than $\pi/4$. The ultimate goal is to determine the neutrino
mixing matrix with at least the same level of precision and redundancy
as the CKM matrix in the quark sector. The size of $\theta_{13}$ plays
a particularly crucial role, since this quantity will set the scale
for the effort necessary to answer the open questions. The need to
determine $\theta_{13}$ has spurred a number of reactor neutrino
experiments~\cite{Anderson:2004pk} using disappearance of $\bar\nu_e$:
Double Chooz~\cite{Ardellier:2004ui}, RENO~\cite{Ahn:2010vy} and Daya
Bay~\cite{Guo:2007ug}. Their discovery reach at $3\,\sigma$ confidence
level will go down to approximately
$\sin^22\theta_{13}=10^{-2}$~\cite{Huber:2009cw,Mezzetto:2010zi}. At
the same time the next generation of long baseline experiments looking
for $\nu_\mu\rightarrow\nu_e$ is underway with T2K~\cite{Itow:2001ee}
and NO$\nu$A~\cite{Ayres:2004js}. Neither T2K nor NO$\nu$A can provide
information on $\deltacp$ beyond a mere indication and even that is
only possible in combination with the data from Daya
Bay~\cite{Huber:2009cw}. The discovery of the mass hierarchy, if
discovery is defined at the usual $5\,\sigma$ confidence level, will
not be possible; even a $3\,\sigma$ evidence is very
unlikely~\cite{Huber:2009cw}. Therefore, it has been widely recognized
that long baseline experiments with physics capabilities far beyond
NO$\nu$A and T2K are necessary, for a review of the possibilities, see
reference~\cite{Bandyopadhyay:2007kx}.

Here, we would like to focus on the superbeam concept, or more
specifically on what is called a wide band beam (WBB). In a superbeam,
muon neutrinos are produced by the decay of pions, where the pions
have been produced by proton irradiation of a solid target. All
neutrino beams relevant in the context of this study use a magnetic
horn to focus and sign-select the pions. In a wide band beam, the
detector is on-axis and thus receives a wide ({\it sic}!)  energy
spectrum of neutrinos. The wide band beam concept makes maximal use of
the available pions and thus provides higher event rates compared to a
narrow band or off-axis beam. The price to pay for the wide spectrum
is that the detector needs to have a very good energy resolution, and
the existence of a high energy tail in the beam will lead to feed down
of neutral current background events. Thus, a wide band beam imposes
unique demands on the detector technology. Apart from allowing for
more events, the wide beam spectrum allows to study a range of $L/E$
values within one experiment, and possibly even to observe more than
only one oscillation maximum\footnote{We will use the term oscillation
  maximum also for disappearance channels, where actually a minimum in
  the survival probability is observed.}. On the level of oscillation
probabilities, the ability to observe two or more cycles of the
oscillation obviously allows to distinguish between otherwise
degenerate solutions. Therefore, the observation of the second
oscillation maximum is considered to play an important role in wide
band beam experiments. The purpose of the present paper is to study in
detail and in a quantitative manner whether the assertion of the role
of the second oscillation maximum based on probabilities remains valid
in a full numerical sensitivity calculation. We also include the case
of non-standard interactions, where one expects similar benefits from
the presence of the second oscillation maximum.

In order to perform a full numerical sensitivity calculation we need
to specify the experimental parameters in great detail and therefore
have to constrain the numerical analysis to a specific setup, which we
model to resemble the Fermilab to DUSEL Long Baseline Neutrino
Experiment (LBNE). However, whenever the specifics of the chosen
experimental setup obscure the underlying physics, we will show
results for sensible variations around our chosen setup. In
section~\ref{sec:frame} we discuss the theoretical framework with
respect to standard and non-standard oscillations. In
section~\ref{sec:methods} we spell out the details of the experimental
setup and describe our analysis techniques.  Section~\ref{sec:results}
will contain our results on both three flavor oscillation and
non-standard interactions and finally in section~\ref{sec:summary} we
will summarize our findings and present our conclusion. In
appendix~\ref{app:alt} we show supplementary results on variations of
the total exposure and baseline.

\section{Framework}
\label{sec:frame}

\subsection{Three flavor oscillation}
\label{sec:3f}

In this paper our main concern is the measurement of the transition
probabilities $P(\nu_\mu\rightarrow\nu_e)$ and
$P(\bar\nu_\mu\rightarrow\bar\nu_e)$. Since the baseline considered is
longer than $1,000\,\mathrm{km}$, matter effects will play an
important role.  While the underlying Hamiltonian describing
oscillations in the presence of a matter potential is quite simple,
the resulting expressions for the exact oscillation probabilities are
not.  Therefore, a plethora of approximations have been devised, for
an overview see reference~\cite{Akhmedov:2004ny}. Even these
approximate solutions have a very rich structure; in particular, for
fixed energy, any given value of the oscillation probability can
typically be realized by different combinations of oscillation parameters.
The possible degenerate solutions can be classified into the intrinsic
ambiguity~\cite{BurguetCastell:2001ez}, the sign of $\dm{31}$
ambiguity~\cite{Minakata:2001qm} and the octant
ambiguity~\cite{Fogli:1996nn}. Combined, these three ambiguities give
rise to what has become known as the eightfold
degeneracy~\cite{Barger:2001yr}.  A recent comprehensive analytical
discussion of the eightfold degeneracy can be found in
reference~\cite{Minakata:2010zn}. In the context of long baseline
experiments the most worrisome degeneracy stems from a combination of
the intrinsic and sign ambiguity which can lead to a phenomenon called
$\pi$-transit~\cite{Huber:2002mx}, in which a CP violating true
solution is mapped into a CP conserving fake solution. A large number
of possible remedies has been proposed in the literature, here we
focus on the proposal to use not only neutrino and anti-neutrino
events from the 1$^\mathrm{st}$ oscillation maximum, but also from the
2$^\mathrm{nd}$ oscillation maximum.

\begin{figure}[t]
  \begin{center}
    \includegraphics[width=0.5\textwidth]{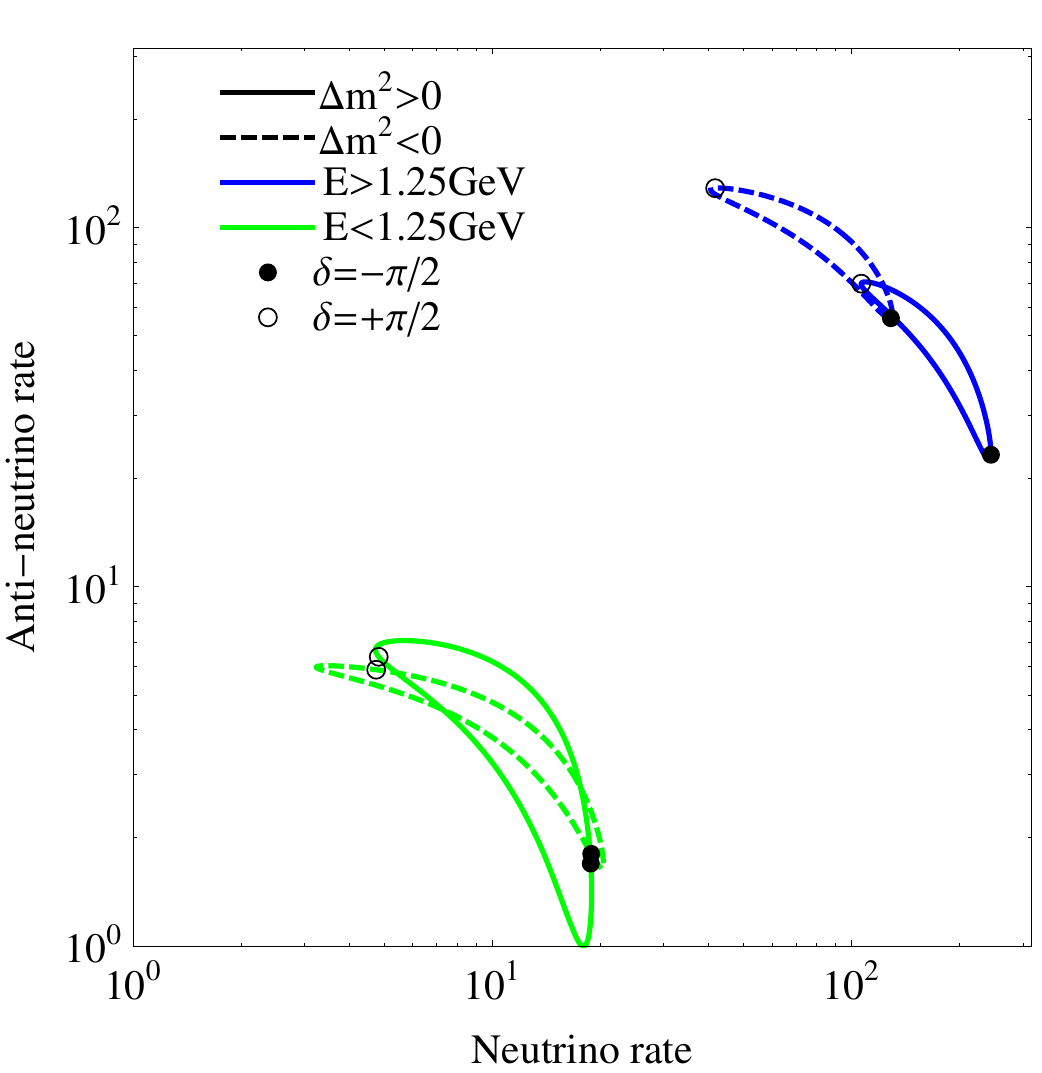}
  \end{center}
  \caption{Bi-rate plot for a typical wide band beam for the
    1$^\mathrm{st}$ (blue/dark gray) and 2$^\mathrm{nd}$ (green/light gray)
    oscillation maximum for normal (solid) and inverted (dashed) hierarchy.
    Along each of the banana-shaped curves, the CP phase $\deltacp$ varies from
    $-\pi$ to $\pi$, where $\deltacp=-\pi/2$ is denoted by a solid disk and
    $\deltacp=+\pi/2$ by an open circle. We have assumed
    $\sin^22\theta_{13}=0.025$, and the other oscillation parameters
    are chosen according to eq.~\eqref{eq:default}.}
  \label{fig:birate}
\end{figure}

In figure~\ref{fig:birate} we illustrate how the use of the
2$^\mathrm{nd}$ oscillation maximum can alleviate the sign degeneracy.
At a baseline of $1,300\,\mathrm{km}$, the first oscillation maximum
occurs with $\Delta m_{31}^2=2.4\times 10^{-3}\,\mathrm{eV}^2$ for
$E_\nu=2.5\,\mathrm{GeV}$ and the second oscillation maximum is at
$E_\nu=0.84\,\mathrm{GeV}$. The first zero of the oscillation term
occurs at $E_\nu=1.25\,\mathrm{GeV}$, and we will use this energy to
separate events from the two oscillation maxima, assigning all events
below to the second maximum and all above to the first. In
figure~\ref{fig:birate} we show so called bi-rate plots. In a bi-rate
plot $\theta_{13}$ is kept fixed and the coordinates are the total
number of events in the neutrino channel and in the anti-neutrino
channel, respectively.  For each possible choice of the CP phase one
obtains a point in this kind of diagram, and as the CP phase is
continuously varied from $-\pi$ to $+\pi$ the points trace out a
banana-shaped curve (or, on a linear scale, an
ellipse)~\cite{Winter:2003ye}. This is similar to a bi-probability
plot~\cite{Minakata:2001qm}, but avoids the problem of choosing a
neutrino energy for plotting and thus allows a closer approximation of
the experimental realities.  In figure~\ref{fig:birate}, we moreover
separated the event sample into events below $1.25\,\mathrm{GeV}$
(green/light gray curves) and events above that energy (blue/dark gray
curves). The solid lines are for normal hierarchy and the dashed ones
for inverted. The solid disks indicate the event rates for
$\deltacp=-\pi/2$, whereas the open circles indicate the event rates
for $\deltacp=+\pi/2$. Focusing on the $2^\mathrm{nd}$ maximum
(green/light gray lines), we see that the ``bananas'' for both
hierarchies are very similar and occupy essentially the same area in
the event rate plane.  Moreover, the event rates at the two maximally
CP violating values of $\deltacp=\pm\pi/2$ do not change when going
from one hierarchy to the other. For the $1^\mathrm{st}$ maximum
(blue/dark gray lines), the two ``bananas'' are very different and the
event rates for the same value of $\deltacp$ change greatly when
switching the hierarchy. Therefore, given enough statistics, we expect
the measurement in the $2^\mathrm{nd}$ maximum to provide a clean
value for $\deltacp$, but no information on the hierarchy. The
measurement in the $1^\mathrm{st}$ maximum, on the other hand, should
yield strong evidence for the mass hierarchy, but may suffer from
degeneracies for the determination of $\deltacp$. The combination of
the two maxima should result in a clear and unambiguous determination
of both the mass hierarchy \emph{and} the CP phase.

\subsection{Non-standard Interactions}
\label{sec:nsi}

At energies of a few GeV, relevant to accelerator neutrino oscillation
experiments, the effects of new physics, which is expected at or above
the electroweak scale, can be parametrized in terms of an effective
theory. Some types of low-scale new physics can also be parametrized
that way~\cite{Joshipura:2003jh,Nelson:2007yq}.
A well known example for the use of effective theory is the
Fermi theory of nuclear beta decay. In this paper, we will use such
non-standard neutrino interactions (NSI) as a benchmark scenario for
deviations from the standard three-flavor oscillation framework, but
we should keep in mind that new physics in the neutrino sector can
also have different manifestations; examples are CPT violation or
mixing between active and sterile neutrinos.  Typical operators
inducing non-standard neutrino interactions (NSI) are
\begin{align}
  \mathcal{L}_{\rm CC} &\supset -2\sqrt{2} G_{F}
    \eps^{{\rm CC},f,f^\prime}_{\alpha\beta}
    \left[ \bar{\nu}_\alpha \gamma^{\rho} P_L \ell_\beta \right]
    \left[ \bar{f} \gamma_{\rho} P_L f^\prime \right] + {\rm h.c.}
  \label{eq:NSI-CC} \\
\intertext{(charged current NSI) and}
  \mathcal{L}_{\rm NC} &\supset -2\sqrt{2} G_{F}
    \eps^{{\rm NC},f}_{\alpha\beta}
    \left[ \bar{\nu}_\alpha \gamma^{\rho} P_L \nu_\beta \right]
    \left[ \bar{f} \gamma_{\rho} P_L f \right] + {\rm h.c.} \,.
  \label{eq:NSI-NC}
\end{align}
(neutral current NSI). Here, $G_F = \sqrt{2} g^2 / 8 M_W^2$ is the
Fermi constant, $P_L = (1 - \gamma^5)/2$, $\alpha$ and $\beta$ are
flavor indices of the neutrinos $\nu$ and the charged leptons $\ell$,
and the fermions $f$ and $f^\prime$ are the members of an arbitrary
weak doublet. The parameters $\eps^{{\rm
    CC},f,f^\prime}_{\alpha\beta}$ and $\eps^{{\rm
    NC},f}_{\alpha\beta}$ give the relative magnitude of the NSI
compared to standard weak interactions. For new physics around the TeV
scale, we expect their absolute values to be of order $10^{-3}$--$10^{-2}$.
In the presence of new degrees of freedom below the electroweak scale,
NSI could be larger and also expectations for the magnitude of NSI
based on effective theory approaches~\cite{Gavela:2008ra,
Antusch:2008tz} may be too conservative. Note that
equations~\ref{eq:NSI-CC} and~\ref{eq:NSI-NC} include only $V-A$ type
interactions, but in principle, more general Lorentz structures are
possible (see e.g.\ refs.~\cite{Kopp:2007ne,Kopp:Phd-Thesis} for an
overview).

For phenomenological purposes, it is convenient to parametrize NSI in
a slightly different way. Consider the $\nu_\alpha \to \nu_\beta$
oscillation probability at baseline $L$,
\begin{align}
  P_{\alpha\beta} = |\bra{\nu_\beta} e^{-i H L} \ket{\nu_\alpha}|^2 \,,
\end{align}
with the Hamiltonian
\begin{align}
  H &= U \begin{pmatrix}
           0 &         & \\
             & \sdm/2E & \\
             &         & \ldm/2E
         \end{pmatrix} U^\dag + V_{\rm MSW} \,,
\end{align}
where $U$ is the leptonic mixing matrix, $E$ is the neutrino energy,
and $V_{\rm MSW}$ is the $3 \times 3$ matrix describing matter
effects. In the presence of CC NSI, the initial and final states get
modified according to
\begin{align}
  \ket{\nu^s_\alpha} = \ket{\nu_\alpha}
    + \sum_{\beta=e,\mu,\tau} \eps^s_{\alpha\beta} \ket{\nu_\beta} \quad\text{and}\quad
  \bra{\nu^d_\beta} = \bra{\nu_\beta}
    + \sum_{\alpha=e,\mu,\tau} \eps^d_{\alpha\beta} \bra{\nu_\alpha} \,,
\end{align}
respectively. The parameters $\eps^s_{\alpha\beta}$ and
$\eps^d_{\alpha\beta}$, which are closely related to the parameters
$\eps^{{\rm CC},f,f^\prime}_{\alpha\beta}$ defined
above~\cite{Kopp:2007ne,Kopp:Phd-Thesis}, describe non-standard
admixtures to the neutrino states produced in association with a
charged lepton of flavor $\alpha$ or detected in a process involving a
charged lepton of flavor $\beta$, respectively.  Note that in
$\eps^s_{\alpha\beta}$, the first index corresponds to the flavor of
the charged lepton, and the second one to that of the neutrino, while
in $\eps^d_{\alpha\beta}$, the order is reversed. The matrices
$(1+\eps^s)$ and $(1+\eps^d)$ need not be unitary, {\it i.e.}\
$\ket{\nu^s_\alpha}$ and $\ket{\nu^d_\alpha}$ are not required to form
complete orthonormal sets of basis vectors in the Hilbert space. Instead
of considering an oscillation probability $P$ normalized to unity, it
is therefore more useful to consider the \emph{apparent} oscillation probability
$\tilde{P}(\nu^s_\alpha \to \nu^d_\beta)$, defined as the number
of neutrinos produced together with a charged lepton of flavor $\alpha$
and converting into a charged lepton of flavor $\beta$ in the detector,
divided by the same number in the absence of oscillations and non-standard
interactions. The apparent oscillation probability is given by
\begin{align}
  \tilde{P}(\nu^s_\alpha \to \nu^d_\beta)
  &= |\bra{\nu^d_\beta} e^{-i \tilde{H} L} \ket{\nu^s_\alpha}|^2 \nonumber\\
  &= \big| (1 + \eps^d)_{\gamma\beta} \, \big( e^{-i \tilde{H} L}
  \big)_{\gamma\delta}
  (1 + \eps^s)_{\alpha\delta} \big|^2           \nonumber\\
  &= \Big| \Big[ \big( 1 + \eps^d \big)^T \,\, e^{-i \tilde{H} L} \,\,
  \big( 1 + \eps^s \big)^T \Big]_{\beta\alpha} \Big|^2 \,,
  \label{eq:P-NSI}
\end{align}
where $\tilde{H} = U \, \diag(0, \sdm/2E, \ldm/2E) \, U^\dag + \tilde{V}_{\rm MSW}$. The
modified matter potential is
\begin{align}
  \tilde{V}_{\rm MSW} = \sqrt{2} G_F N_e
  \begin{pmatrix}
    1 + \eps^m_{ee}       & \eps^m_{e\mu}       & \eps^m_{e\tau}  \\
        \eps^{m*}_{e\mu}  & \eps^m_{\mu\mu}     & \eps^m_{\mu\tau} \\
        \eps^{m*}_{e\tau} & \eps^{m*}_{\mu\tau} & \eps^m_{\tau\tau}
  \end{pmatrix} \,,
\end{align}
with $\eps^m_{\alpha\beta}$ being closely related to the $\eps^{{\rm
NC},f}_{\alpha\beta}$ from equation~\ref{eq:NSI-NC}. As explained
above, the magnitude of the $\eps^{s,d,m}$ parameters is expected to
be at or below the $10^{-2}$ level for new physics at the TeV scale.

Model-independent experimental bounds on the $\eps^{s,d,m}$ parameters are
typically of $\mathcal{O}(10^{-2} - 1)$~\cite{Davidson:2003ha,
GonzalezGarcia:2007ib, Biggio:2009nt}. In a specific model, however, the
bounds may be much stronger because in most models neutrino NSI are
accompanied by charged lepton flavor violation, which is strongly constrained
by precision tests of the electroweak theory and by rare decay searches.  From
a model building point of view, it is therefore not easy to realize large
non-standard neutrino interactions that can saturate the experimental
bounds~\cite{Gavela:2008ra, Antusch:2008tz}.
 
Obviously, with any new experiment, we need to compare the expected
bounds on new physics with the ones we already have. We use the bounds
derived in~\cite{Davidson:2003ha,GonzalezGarcia:2007ib,Biggio:2009nt}
as our benchmark. In particular, we use the 90\% confidence level
constraints $|\eps^m_{ee}| < 4.2$, $|\eps^m_{e\mu}| < 0.33$, $|\eps^m_{e\tau}| < 3.0$
$|\eps^m_{\mu\mu}| < 0.068$, $|\eps^m_{\mu\tau}| < 0.063$, $|\eps^m_{\tau\tau}| < 0.2$.
Note that the relatively strong constraints on $|\eps^m_{\mu\tau}|$ and
$|\eps^m_{\tau\tau}|$ have been derived from atmospheric neutrino data in a two-flavor
framework; when three-flavor effects---in particular correlations
between different types of NSI---are taken into account, these
bounds may become somewhat weaker~\cite{Blennow:2008ym, Friedland:2005vy}.

\section{Methods}
\label{sec:methods}

\subsection{Experimental setup}
\label{sec:setup}

To assess the sensitivity of a wide band neutrino beam to standard and
non-standard oscillation physics, we have performed simulations using
the {\sf GLoBES} software~\cite{Huber:2004ka,Huber:2007ji}, with an implementation
of NSI developed in refs.~\cite{Kopp:2007mi,Kopp:2007ne,Kopp:2006wp}. Our experiment
description follows the LBNE proposal for a long-baseline neutrino
beam from Fermilab to DUSEL, but our results will hold qualitatively
also for other wide band beam experiments.

\subsubsection{Beam}

For the neutrino beam, we consider the options listed in
table~\ref{tab:beams}. We use the unit protons on target (pot) since
it is the usual measure of integrated luminosity, $\mathcal{L}$, for
this kind of experiments. To compare this with other beams of
different energy it is useful to convert this result to equivalent
beam power, $P$, where we assume that the beam is on for
$2 \times 10^7\,\mathrm{s}$ per tropical year.
$$
P=0.801\,\left(\frac{E_p}{\mathrm{GeV}}\right)\left(\frac{\mathcal{L}}{10^{20}\,\text{pot per year}}\right)\,\mathrm{kW}\,.
$$
With this in mind, the luminosities given in table~\ref{tab:beams}
correspond to about 6 years of running (3 years in neutrino mode + 3
years in anti-neutrino mode) at a beam power of either $\sim 700\,\mathrm{kW}$
or $\sim 2,300\,\mathrm{kW}$. We have also studied the performance of
a $60\,\mathrm{GeV}$ beam, which would have the advantage of lower
backgrounds, at the expense of less statistics. Since we found only
very minor performance differences between the $60\,\mathrm{GeV}$ and
$120\,\mathrm{GeV}$ options, we restrict the discussion to the
$120\,\mathrm{GeV}$ beam in the following.  Simulated spectra for all beam
options have been kindly provided to us by the LBNE
collaboration~\cite{Bishai:PrivComm}.
\begin{table}
  \centering
  \begin{ruledtabular}
  \begin{tabular}{rrl}
    Proton energy & pot per polarity     & Comment \\\hline
    120~GeV     &  $22 \times 10^{20}$ & accelerator complex without Project X \\
    120~GeV     &  $72 \times 10^{20}$ & accelerator complex with Project X  \\
  \end{tabular}
  \end{ruledtabular}
  \caption{Neutrino beam configurations considered in our simulations.}
  \label{tab:beams}
\end{table}

\subsubsection{Detectors}

We assume the far detector to be located at a baseline $L =
1,300\,\mathrm{km}$, corresponding to the distance from Fermilab to DUSEL,
which translates into an energy $E\sim2.5\,\mathrm{GeV}$ for the
1$^\mathrm{st}$ oscillation maximum. In order to be sensitive to both
oscillation maxima, and given the beam spectra, the detector needs to
have good efficiency in the energy range from $0.5-4\,\mathrm{GeV}$.
The number of events in the 1$^\mathrm{st}$ oscillation maximum will
be significantly larger than that in the 2$^\mathrm{nd}$ maximum.
Currently, two detector technologies are considered in this context

\begin{enumerate}

\item A water \v{C}erenkov (WC) detector with a fiducial mass of
  $200\,\mathrm{kt}$. As demonstrated by Super-Kamiokande, this type
  of detector allows for a clean separation of muon and electron
  quasi-elastic events. However, its application at GeV energies
  requires careful consideration of possible backgrounds from neutral
  current events giving rise to energetic neutral pions. A
  considerable amount of work went into studying this
  issue~\cite{Yanagisawa:2007zz,Dufour:2010vr}. Both studies agree
  quite well and we use the results from
  reference~\cite{Yanagisawa:2007zz}. The {\sf GLoBES} description of
  this WC detector is based on
  references~\cite{Barger:2006vy,Diwan:2006qf}. This simulation
  includes energy-dependent efficiency tables, smearing matrices, and
  background estimates based on Monte-Carlo codes developed by the
  Super-Kamiokande collaboration~\cite{Yanagisawa:2007zz}.  We include
  both events from the $\nu_e$ appearance channel as well as the
  $\nu_\mu$ disappearance channel, for both neutrino and anti-neutrino
  running.  

\item A liquid argon (LAr) time projection chamber (TPC) with a
  fiducial mass of $34\,\mathrm{kt}$. For the case of LAr, only much
  less detailed and accurate simulations are available since no large
  LAr detector has ever been operated. The LAr description is based on
  references~\cite{Fleming:PrivComm, Barger:2006kp}.  We include the
  $\nu_e$ and $\bar\nu_e$ appearance channels, with backgrounds from
  the intrinsic $\nu_e$/$\bar{\nu}_e$ contamination of the beam,
  misidentified muons, and neutral current events. Our background
  estimate is conservative because the very high spatial resolution
  and the ability to detect very low energy particles in a LAr
  detector might allow for a much more efficient rejection of neutral
  current events.  For the $\nu_\mu$ ($\bar\nu_\mu$) disappearance
  channel, the main backgrounds stem from neutral current events (we
  assume a rejection efficiency of 99.5\%) and from the
  $\bar{\nu}_\mu$ ($\nu_\mu$) ``wrong sign'' contamination of the
  beam. Since the oscillation probabilities in the $\nu_\mu$ and
  $\bar{\nu}_\mu$ disappearance channels are similar (except for the
  sub-leading contribution from matter effects), the latter background
  does not constitute a problem.
\end{enumerate}

For both detectors, the neutrino cross sections are based on
\cite{Messier:1999kj,Paschos:2001np}; they are computed for water and
isoscalar targets, respectively. In an actual experiment great care
needs to be taken to correctly model the cross sections, including
nuclear effects. In our case, since we are using the same cross
section to compute the data and perform the fit to that simulated
data, any error due to the omission of nuclear effects will cancel.
 
In our simulations we also include a near detector; for standard
oscillation physics, its main effect is to reduce systematic
uncertainties in the far detector, but for charged current NSI
searches, it is valuable also as a standalone detector and its
inclusion in the simulation is imperative. Since no specific
technology has been chosen for the LBNE near detector(s) yet, we take
a generic approach and assume the near detector to have identical
properties (resolution, efficiencies, backgrounds, etc.) as the far
detector, but a fiducial mass of only 1~kt. Furthermore, we assume the
geometric acceptance of the near and far detectors to be the same, which
greatly simplifies the calculation, but is very difficult to achieve in
practice. In reality, the optimum choice of near detector technology and
geometry may be very different for the WC and LAr cases, and thus also the
effective systematic uncertainties may be quite different.

\begin{figure}[t]
  \begin{center}
    \includegraphics[width=\textwidth]{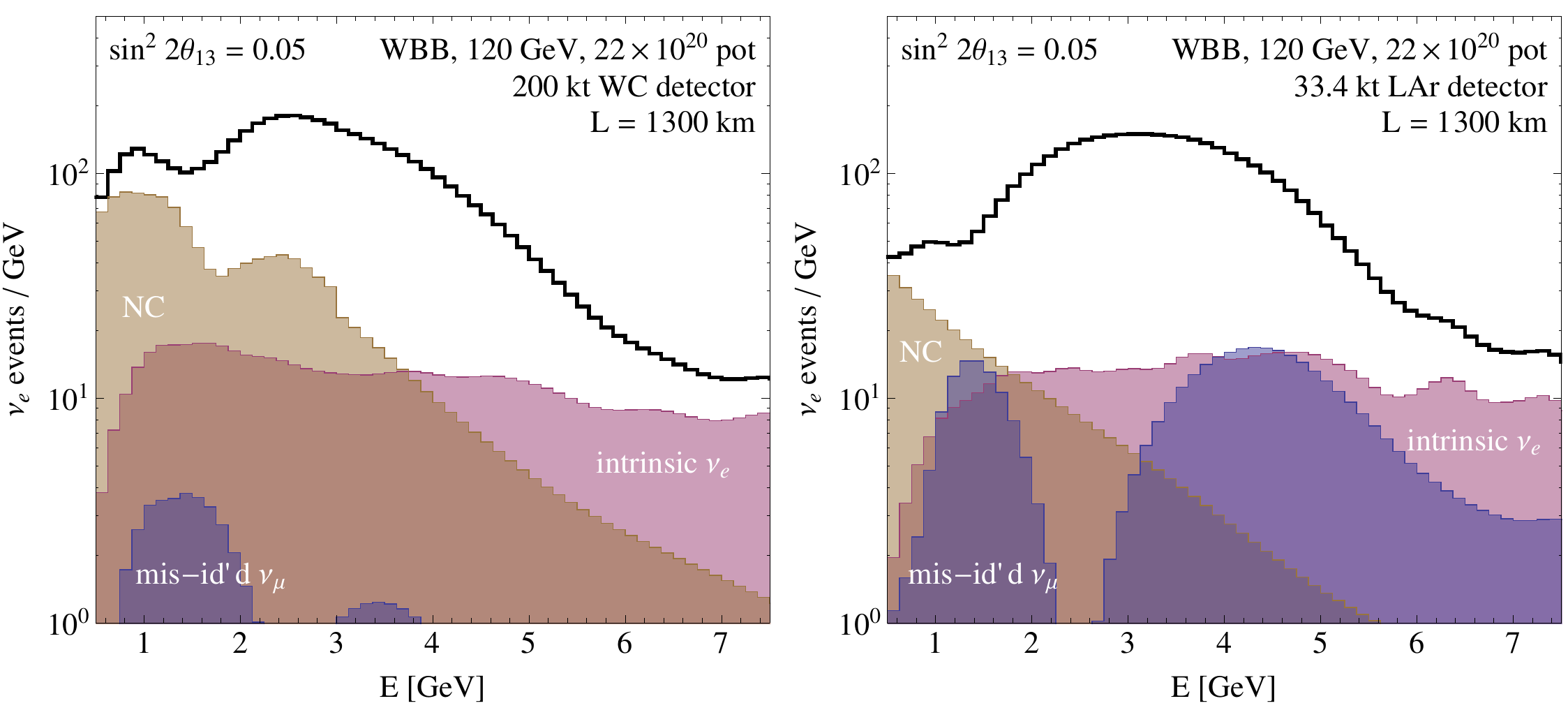}
  \end{center}
  \caption{Expected event rates in the $\nu_e$ appearance channel for
    $\sin^2 2\theta_{13} = 0.05$, $\deltacp = 0$ as a function of the
    reconstructed neutrino energy. The black histograms show the signal +
    background rates, while the filled red, blue, and beige histograms depict
    the backgrounds due to the intrinsic $\nu_e$ contamination of the beam,
    misidentified $\nu_\mu$ events, and neutral current events, respectively.
    Except for $\theta_{13}$ and $\deltacp$, the oscillation parameters are
    chosen according to eq.~\eqref{eq:default}.}
  \label{fig:raw-rates}
\end{figure}

For illustration, we show in fig.~\ref{fig:raw-rates} the expected
event rates in the $\nu_e$ appearance channel for both detectors. It
is clear that for small $\theta_{13}$, backgrounds will be a
limitation.  In particular, neutral current events contaminate the
second oscillation peak.

\subsection{Analysis}

To analyze the simulated data sets and to compute exclusion limits and
allowed parameter regions, we use a $\chi^2$ analysis following
ref.~\cite{Huber:2002mx}.  Our $\chi^2$ function has the form
\begin{align}
  \chi^2 = \min_{\vec{a}} \Big[ \,
    \sum_{d = N, F} \; \sum_{s=\nu_e, \bar{\nu}_e, \nu_\mu, \bar{\nu}_\mu}
    \sum_{j=1}^{\# \text{ of bins}}
    \chi^2_{\rm stat}\Big( N^{\rm obs}_{d,s,j}, N^{\rm th}_{d,s,j}(\vec{\Theta}), \vec{a} \Big)
  + \sum_i \frac{a_i^2}{(\sigma^a_i)^2} \Big] 
  + \sum_j \frac{(\Theta_j - \Theta^{(0)}_j)^2}{(\sigma^\Theta_j)^2} \,,
  \label{eq:chi2}
\end{align}
where $N^{\rm obs}_{d,s,j}$ and $N^{\rm th}_{d,s,j}(\vec{\Theta})$ are
the observed and theoretically predicted event rates for detector $d$
($d = N$ (near) or $F$ (far)), event sample $s$ ($s = \nu_e,
\bar{\nu}_e, \nu_\mu, \bar{\nu}_\mu$), and bin $j$. The vector
$\vec{\Theta}$ stands for the oscillation parameters, while $\vec{a}$
contains the systematical biases. The first term on the right hand
side of equation~\ref{eq:chi2} is the statistical contribution to
$\chi^2$, while the second term contains pull terms that disfavor
values of the biases $a_i$ much larger than the associated systematic
uncertainties $\sigma^a_i$. In a similar way, the last term of
equation~\ref{eq:chi2} is used to confine the oscillation parameters to
within the region determined by other experiments, where, for each
oscillation parameter $\Theta_j$, $\Theta^{(0)}_j$ denotes the
externally given best fit value and $\sigma^\Theta_j$ the $1\sigma$
uncertainty on that value. In our simulations, we include such
external prior terms only for the solar oscillation parameters
$\theta_{12}$ and $\Delta m_{21}^2$ to which the wide band beam is not
sensitive. We assume the solar parameters to be known to within 5\% at
the $1\sigma$ level, while for all other oscillation parameters, we
set $\sigma^\Theta_j = \infty$. The default oscillation parameters
used in this study are, in agreement with current
fits~\cite{Maltoni:2004ei},
\begin{eqnarray}
\label{eq:default}
  \sin^2\theta_{12}=0.32\,, &\quad& \theta_{13} = 0\,,\nonumber\\
  \theta_{23} = \frac{\pi}{4}\,, &\quad& \deltacp = \frac{3}{2}\pi\,,\nonumber \\
  \Delta m^2_{21} = +7.6\times10^{-5}\,\mathrm{eV}^2\,, &\quad&
  \Delta m^2_{31} =+ 2.4\times10^{-3}\,\mathrm{eV}^2\,,
\end{eqnarray}
and we use a conservative 5\% uncertainty on the matter density.

The systematic errors we include in our study are listed in
Table~\ref{tab:sys} for both detector technologies. We treat the
normalization of the beam flux and of the background contributions to
the $\nu_e$ and $\bar{\nu}_e$ event samples as completely free
parameters, i.e.\ we do not include pull terms for them. Moreover,
we allow for uncorrelated systematic biases in the number of signal
and background events in each event sample and each detector.
Systematic uncertainties are assumed to be completely uncorrelated
between the neutrino and anti-neutrino runs of the experiment.
\begin{table}
  \centering
  \begin{ruledtabular}
  \begin{tabular}{lccc}
    &WC       &LAr     &N/F correlated?\\
    \hline
    Beam flux [\%]                                        &$\infty$ &$\infty$& yes \\
    Intrinsic background [\%]                             &$\infty$ &$\infty$& yes \\
    Signal normalization for $\nu_e$ sample [\%]          &0.7      &0.7     & no  \\
    Background normalization for $\nu_e$ sample [\%]      &3.5      &7.0     & no  \\
    Signal normalization for $\nu_\mu$ sample [\%]        &0.7      &3.5     & no  \\
    Background normalization for $\nu_\mu$ sample [\%]    &7.0      &7.0     & no  \\
  \end{tabular}
  \end{ruledtabular}
  \caption{Systematic uncertainties assumed in our simulations. All systematic
    errors are assumed to be completely uncorrelated between the neutrino and
    anti-neutrino runs of the experiment. Note that uncertainties that are
    uncorrelated between the near (N) and far (F) detectors will add in
    quadrature when translated into an error on the measured oscillation
    probability.  For example the near-far uncorrelated 0.7\% uncertainty in the
    number of $\nu_e$ signal events would translate into a 1\% uncertainty on the
    measured oscillation probability.}
  \label{tab:sys}
\end{table}

We use the following performance indicators to estimate the
sensitivity of the experiment to standard oscillation physics
\begin{itemize}
\item {\bf $\theta_{13}$ discovery reach.} For each combination of
  true $\theta_{13}$ and true $\deltacp$, we compute the
  expected experimental event rates, and then perform a $\chi^2$ fit
  assuming a test value of $\theta_{13} = 0$. If, for a particular
  combination of $\theta_{13}^{\rm true}$ and $\deltacp^{\rm true}$,
  the fit disagrees with the simulated data at a given confidence
  level, we say that that this $\theta_{13}^{\rm true}$ and
  $\deltacp^{\rm true}$ are within the discovery reach of the
  experiment at that confidence level. In the fit, we marginalize over
  all oscillation parameters except $\theta_{13}$ (which is kept fixed
  at zero) as well as the matter density.

\item {\bf Discovery reach for the normal mass hierarchy (NH)} For
  each point in the $\theta_{13}^{\rm true}$--$\deltacp^{\rm true}$
  plane, we simulate the event rates assuming a normal mass hierarchy,
  and then attempt a fit to the simulated data assuming the inverted
  hierarchy. If the fit is incompatible with the data at a given
  confidence level, we say that the chosen combination of
  $\theta_{13}^{\rm true}$ and $\deltacp^{\rm true}$ is within the NH
  discovery reach of the experiment.

\item {\bf CP violation (CPV) discovery reach.} For each point in the
  $\theta_{13}^{\rm true}$--$\deltacp^{\rm true}$ plane, we simulate
  the expected event rates and then attempt fits assuming $\deltacp =
  0$ and $\deltacp = \pi$.  If the fits are able to exclude the CP
  conserving solutions at a given confidence level, we say that the
  chosen combination of $\theta_{13}^{\rm true}$ and $\deltacp^{\rm
    true}$ is within the CPV discovery reach of the experiment.

\item {\bf Sensitivity to the octant of $\boldsymbol{\theta_{23}}$.} For each point
  in the $\theta_{23}^{\rm true}$--$\theta_{13}^{\rm true}$ plane, we
  simulate the expected event rates and then attempt a fit in which
  all parameters are marginalized over, but $\theta_{23}$ is forced
  to lie in the ``wrong'' octant, \emph{i.e.}\ between $\pi/4$ and
  $\pi/2 - \theta_{23}^{\rm true}$. If the fit is
  incompatible with the simulated data at a given confidence level, we
  say that the experiment is sensitive to the octant of $\theta_{23}$
  at that confidence level.
\end{itemize}

When discussing non-standard neutrino interactions, we will use the
{\bf NSI discovery reach} as a performance indicator, which we define
in analogy to the $\theta_{13}$ discovery reach: For each set of
true NSI parameters, we check whether a standard oscillation fit
neglecting NSI is compatible with the data at a given confidence
level. If this is not the case, the chosen NSI parameters are within
the experimental discovery reach.

\section{Results}
\label{sec:results}

\subsection{Standard oscillation}

\begin{figure}
  \begin{center}
    \includegraphics[width=\textwidth]{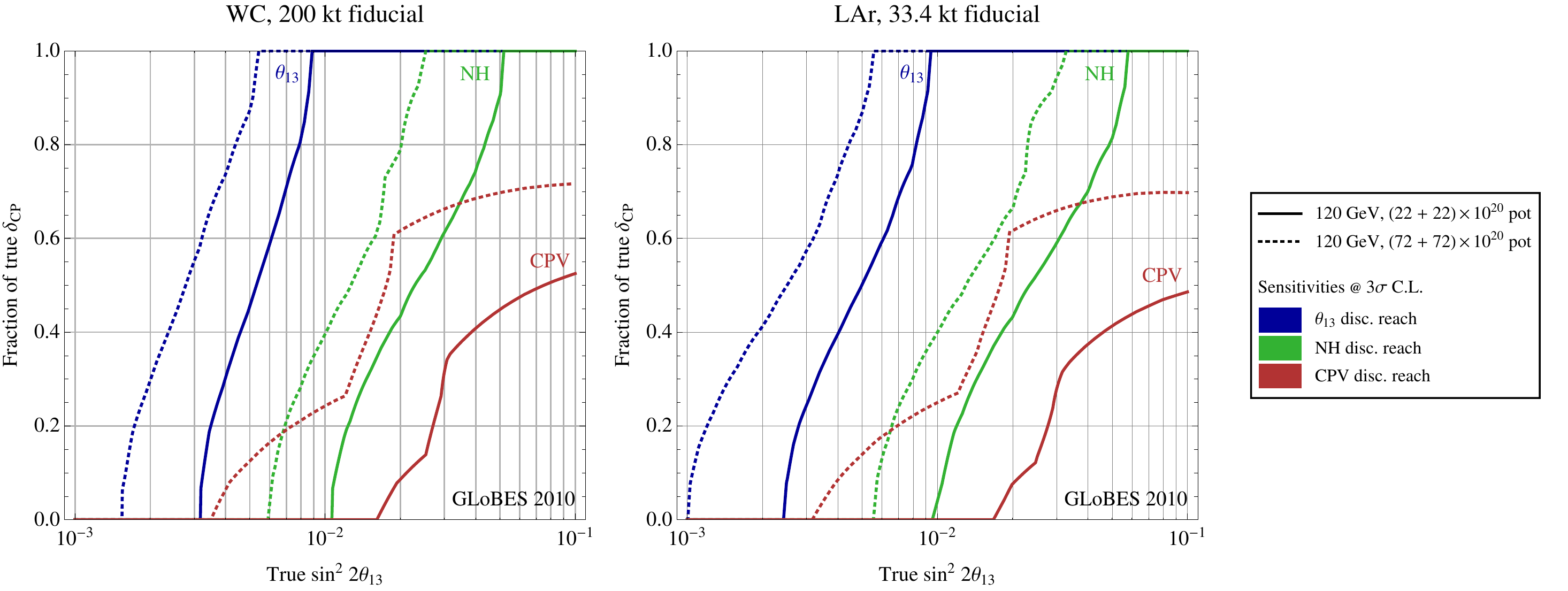}
  \end{center}
  \caption{Sensitivity to standard oscillation physics in a wide band
    beam for a $200\,\mathrm{kt}$ (fiducial) water \v{C}erenkov
    detector (left) and a $34\,\mathrm{kt}$ (fiducial) liquid argon
    detector (right). The results are shown at $3\,\sigma$ confidence
    level.}
  \label{fig:wc-vs-lar}
\end{figure}

First, we summarize the physics performance of the default
setup, as defined in section~\ref{sec:setup}, with respect to three
flavor oscillation. Figure~\ref{fig:wc-vs-lar} depicts the discovery
reaches for CPV, $\theta_{13}$ and the mass hierarchy. In the left
hand panel we show results for a $200\,\mathrm{kt}$ water \v{C}erenkov
detector (WC), whereas in the right hand panel we show the corresponding
results for a $34\,\mathrm{kt}$ liquid argon detector (LAr). We have
checked that the difference in performance between the
$60\,\mathrm{GeV}$ and $120\,\mathrm{GeV}$ proton beams, at equivalent
power, is very small, and therefore we only show the result for
the $120\,\mathrm{GeV}$ beam.
It is apparent from this figure that the performance of the two
detectors, despite a factor of 6 difference in fiducial masses, is
quite similar. If $\sin^22\theta_{13}<0.04$ the beam upgrade provided
by Project X, whose results are shown as dotted lines, is a necessity
to ensure a mass hierarchy determination and to maintain a better than
50\% coverage for CP violation. Even for the largest possible values
of $\theta_{13}$ the CP sensitivity would greatly benefit from a
luminosity upgrade, as also can be seen from
figure~\ref{fig:exposure}.  We have also evaluated the relative precision on
$\sin^22\theta_{13}$, defined by $(\sin^22\theta_{13}^{\rm max}
- \sin^22\theta_{13}^{\rm min}) / \sin^22\theta_{13}^{\rm true}$,
where $\theta_{13}^{\rm min}$ and $\theta_{13}^{\rm max}$ denote the
lower and upper bounds on $\theta_{13}$ that can be expected for a
particular $\theta_{13}^{\rm true}$. We find for
$\sin^22\theta_{13}^{\rm true}=0.1$ that the WC detector measures
$\sin^22\theta_{13}$ with a relative $3\sigma$ error between 33\% and
39\%, depending on the true value of $\deltacp$, while the LAr
detector can achieve a precision between 36\% and 42\%. This result
should be compared with the accuracy obtainable from reactor neutrino
experiments like Daya Bay, which will provide a relative error of 18\%
at $3\sigma$~C.L.~\cite{Huber:2009cw}.
\begin{figure}
  \begin{center}
    \includegraphics[width=0.5\textwidth]{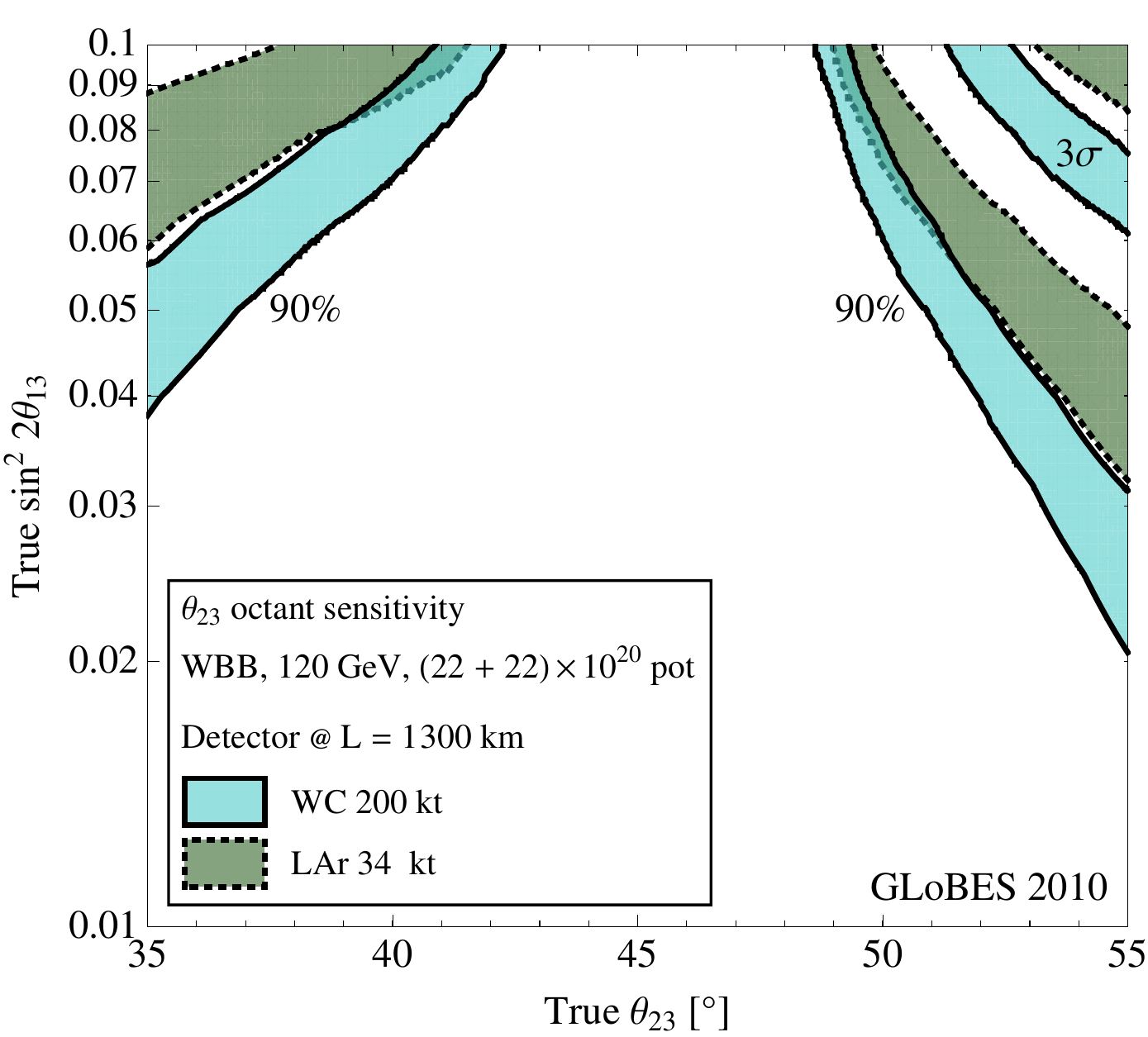}
  \end{center}
  \caption{\label{fig:th23oct} Ability to determine the octant of
    $\theta_{23}$ at $3\,\sigma$ and $90\%$ confidence level for
    $200\,\mathrm{kt}$ WC and and $34\,\mathrm{kt}$ LAr detectors as
    labeled in the legend. The result is shown as a function of
    $\theta_{23}^{\rm true}$ and $\sin^22\theta_{13}^{\rm true}$.
    The width of each region is due to the unknown CP phase.}
\end{figure}
In figure~\ref{fig:th23oct} we study the ability to determine the
octant of $\theta_{23}$ in a WC or LAr detector.
The width of the colored regions in the plot is due to
the marginalization over the unknown CP phase $\deltaCP^{\rm true}$. For this measurement,
we see again that the differences in performance between a WC detector
and a six times smaller LAr detector are small. Determining the
octant of $\theta_{23}$ is a difficult measurement for any experiment.
In particular, for $\theta_{23}$ close to $45^\circ$ this measurement is only
possible for large $\theta_{13}$. The asymmetry in sensitivity between $\theta_{23}^{\rm true}$
above or below $45^\circ$ is due to the partial cancellation between the
octant sensitive terms in $P_{\mu e}$ and matter effects. Therefore,
the sign of the asymmetry will change if we were to change the assumed
true mass hierarchy from normal to inverted in our calculation.

\subsection{2$^\mathrm{nd}$ maximum}

\begin{figure}[t!]
  \begin{center}
    \includegraphics[width=\textwidth]{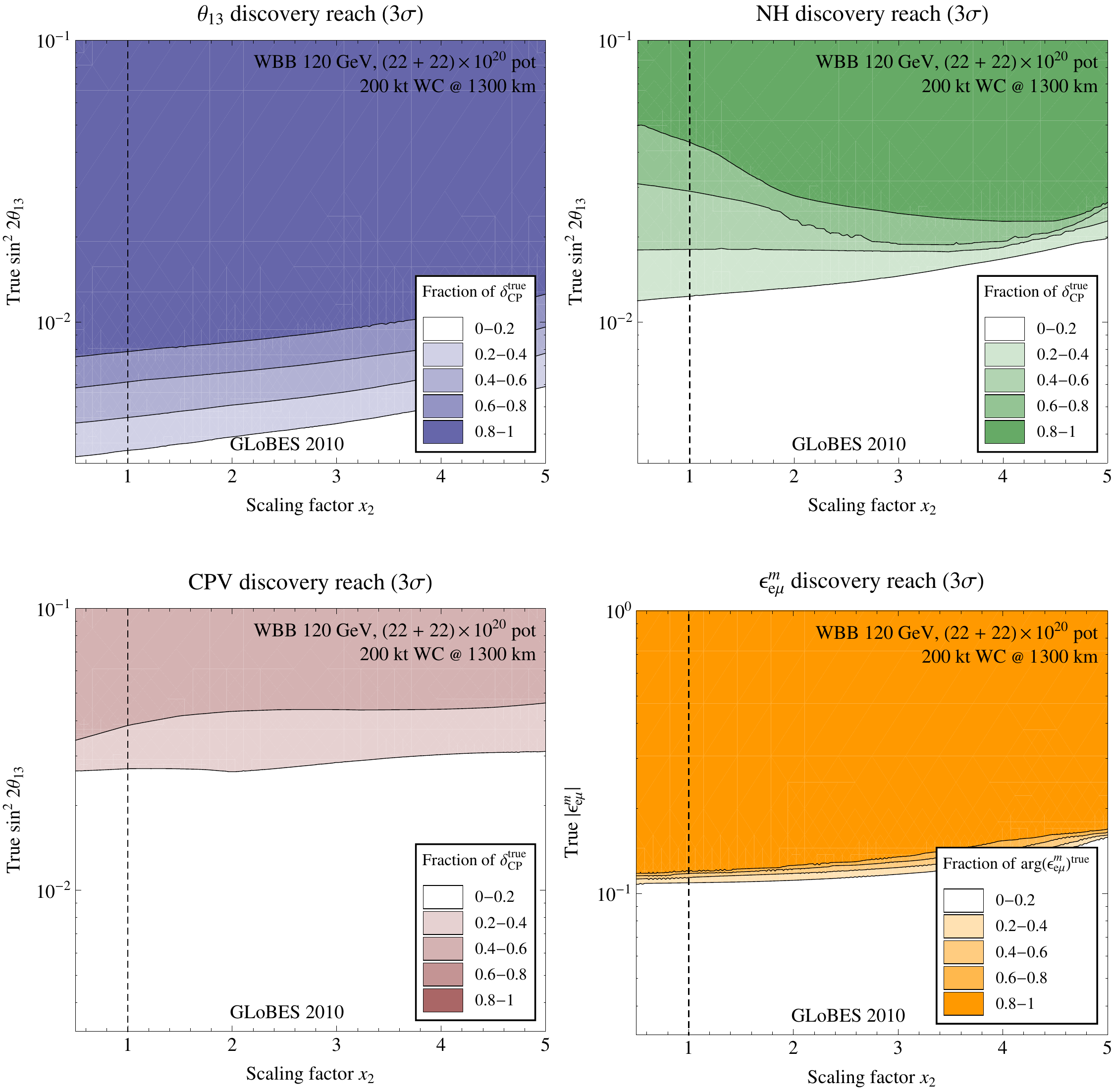}
  \end{center}
  \caption{\label{fig:variation} The dependence of various
    sensitivities, as labeled in the legend of each panel, on the
    scaling parameter $x_2$ as defined in equation~\ref{eq:param}. Shown
    are lines of constant CP fraction in the $x_2$--$\sin^22\theta_{13}^{\rm true}$
    resp.\ $x_2$--$|(\eps^m_{e\mu})^{\rm true}|$ plane. The results are given at
    $3\,\sigma$ confidence level.}
\end{figure}

Now that we have established the baseline performance, we can turn our
attention to the central question of this paper: what is the
\emph{quantitative} impact of data from the $2^\mathrm{nd}$ oscillation
maximum on the physics sensitivities? This question specifically
neglects the issue of how the robustness of an experiment with respect
to unforeseen systematical effects improves due to the data from
$2^\mathrm{nd}$ oscillation maximum. However, the current analysis
does include known systematic effects like normalization errors of
backgrounds and signal. Obviously, if the data from the
$2^\mathrm{nd}$ oscillation maximum can be collected at no or only
very small cost, we are well advised in using it, even if only to
check whether our assumptions about the performance of the experiment
and the underlying physics model are correct. However, in case that
obtaining this data turns out to be costly, we need to understand in
a quantitative way how much one would lose by not having it.

For the baseline setup discussed in the previous section, we can show
that for all standard oscillation measurements, with the exception of
the sensitivity to the mass hierarchy, there is virtually no
difference between an analysis which includes both maxima and one where
we ignore all data with energies below $1.25\,\mathrm{GeV}$. For the
mass hierarchy measurement, the improvement happens for those values
of the CP phase where the $\pi$-transit phenomenon (see
section~\ref{sec:3f}) would strongly reduce the sensitivity. Even
there, the data from the $2^\mathrm{nd}$ maximum is statistically not
significant enough to improve the sensitivity to the level it would
have if there were no $\pi$-transit. 

This result does not imply that the $2^\mathrm{nd}$ maximum makes no
quantitative difference at all, it just shows that, with the specific
experimental setup chosen, the data sample in the $2^\mathrm{nd}$
maximum is too small and the backgrounds are too large (see
figure~\ref{fig:raw-rates}) in order for that data to make a sizable
contribution to the overall $\chi^2$. Therefore, we will now study how
the sensitivities change if there are more events in the energy region
below 1.25~GeV. If we just were to scale up the number of events in
the $2^\mathrm{nd}$ maximum, obviously, we always would find that the
$\chi^2$ becomes larger, since it is a monotonically increasing function
of the total number of events. However, at constant beam power, the
number of pions produced in the target is constant as well.
Therefore, any beam optimization is equivalent to selecting a
different subset of pions leading to different neutrino spectra.
Assuming, furthermore, that the acceptance of the horn and beam pipe
are finite and fixed, any optimization is just a reshuffling of pions
and hence neutrinos of different energies, with the total number of
neutrinos remaining fixed. This inspires the following parametrization:
Let $\phi$ be the total flux of $\nu_\mu$ in the beam, and let
$\phi_1$ ($\phi_2$) be the partial flux in the energy window above
(below) 1.25~GeV, corresponding to the $1^\mathrm{st}$
($2^\mathrm{nd}$) oscillation maximum.  To assess the importance of
the second maximum to the experimental sensitivity, we vary the
fraction of neutrinos below 1.25~GeV, while keeping the total flux
constant.  More specifically, we scale $\phi_2$ with an efficiency
factor $x_2$, and $\phi_1$ with an efficiency factor $x_1 = [\phi_1 +
(1-x_2) \phi_2] / \phi_1$, so that $\phi_1 + \phi_2$ does not change.
$x_2$ can take values between $0$ and $\phi / \phi_2$, with $x_2 = 1$
corresponding to the setup defined in section~\ref{sec:setup}.  For each fixed
$x_2$, we compute $x_1$ separately for the neutrino beam and the anti-neutrino
beam. To summarize this parametrization, we have
\begin{equation}
\label{eq:param}
\phi\equiv \phi_1+\phi_2\quad\mathrm{and}\quad \phi_2\rightarrow x_2\, \phi_2
\quad\mathrm{and}\quad \phi_1\rightarrow \phi-x_2\,\phi_2 
\quad\mathrm{and}\quad x_2\in\left[0,\frac{\phi}{\phi_2}\right]\,.
\end{equation}
In the first three panels of figure~\ref{fig:variation}, we show the
sensitivity to standard oscillation physics as a function of $x_2$. We
see that for all performance indicators except the mass hierarchy, the
optimum occurs at $x_2 < 1$, which implies that we rather have more
events in the $1^\mathrm{st}$ maximum instead of sharing them with the
$2^\mathrm{nd}$ maximum.
\begin{figure}[t!]
  \begin{center}
    \includegraphics[width=\textwidth]{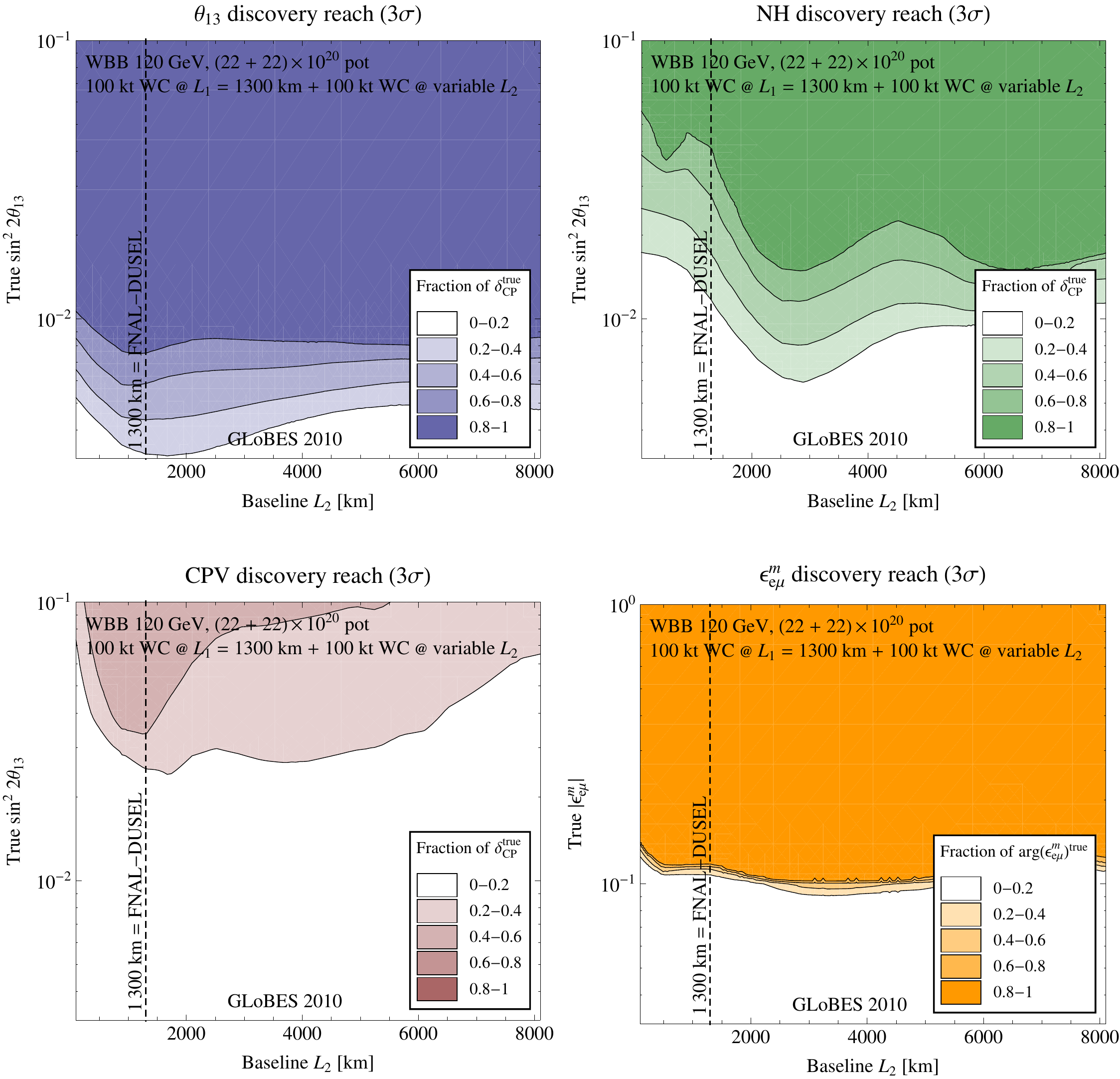}
  \end{center}
  \caption{\label{fig:two-det} The dependence of various
    sensitivities, as labeled in the legend of each panel, on the
    baseline $L_2$ of a second detector. Shown are lines of constant CP
    fraction in the $L_2$--$\sin^2\theta_{13}^{\rm true}$ resp.\
    $L_2$--$|(\eps^m_{e\mu})^{\rm true}|$ plane. The results are
    given at $3\,\sigma$ confidence level.}
\end{figure}

An altogether different way to access events in the $2^\mathrm{nd}$
oscillation maximum is to use a second detector at a longer baseline.
If the two detectors are to be in the same beam and the first one is on-axis,
the second one has to be off-axis due to the curvature of the Earth's
surface. Alternatively, one can also imagine scenarios in which
both detectors are off-axis, either with identical off-axis angles
or with different ones. For example, in the T2KK
setup~\cite{Ishitsuka:2005qi,Dufour:2010vr}, it has been proposed to use
Super-Kamiokande (or a larger Water \v{C}erenkov detector at the same site)
as the detector sensitive to the $1^\mathrm{st}$ maximum, and
supplement it with a second Water \v{C}erenkov detector at a baseline of about
$1,000\,\mathrm{km}$ on the east coast of Korea.  Another
proposal~\cite{Badertscher:2008bp} puts a liquid Argon detector at
about $600\,\mathrm{km}$ on the island of Okinoshima. In both cases,
due to the different off-axis angles, the second detector will be
predominantly sensitive to the $2^\mathrm{nd}$ maximum. A superficial
comparison of the obtainable sensitivities indicates a similar physics
performance, where most of the differences is attributable to the
different overall exposure~\cite{Barger:2006kp}.  In order to allow
for a direct comparison with the results derived in this paper, we
refrain from comparing these setups in detail and study instead the
effects of the addition of a second baseline to the setup we have
introduced in section~\ref{sec:setup}. Since we are interested in the
question of the \emph{general} impact the $2^\mathrm{nd}$ maximum can
have, we will neglect the actual geometry and assume that we have two
identical beams, which allows us to put the second detector on-axis
into this second beam. This is clearly an unrealistic and overly
optimistic assumption. It amounts to doubling the number of protons on
target and in contrast to the proposals centered around
Super-Kamiokande, which all exploit a single beam, leads to higher
event rates in the $2^\mathrm{nd}$ maximum due to it being accessed in
an on-axis beam. However, it allows us to estimate the maximum effect
that events from the $2^\mathrm{nd}$ maximum could have under ideal
circumstances. In other words, if we do not observe an overwhelming
increase in performance under these most favorable conditions, then we
can safely conclude that the $2^\mathrm{nd}$ oscillation maximum,
despite its theoretical merits, in practice is not useful in a
superbeam experiment. The results of this analysis are shown in
figure~\ref{fig:two-det}. For the measurement of $\theta_{13}$ and CP
violation, the performance optimum occurs for a detector location very
close to $1,300\,\mathrm{km}$, which is the position of the first
detector. The sensitivity to the mass hierarchy shows a strong
preference of baselines around $2,500\,\mathrm{km}$, but we remark that
a similar effect is also
seen with only one detector, see figure~\ref{fig:baseline} and also
reference~\cite{Barger:2006vy}. Interestingly, at this distance the
first oscillation minimum is at the peak of the beam flux and thus
both the $1^\mathrm{st}$ and $2^\mathrm{nd}$ maximum contribute about
equally to the rate. 

A further question is whether information from the $2^\mathrm{nd}$
maximum can help in determining the octant of $\theta_{23}$. To this
end we computed the sensitivity to the octant in the same way as shown
in figure~\ref{fig:th23oct} but constraining the data to the
$1^\mathrm{st}$ maximum only. The results are identical to the one in
figure~\ref{fig:th23oct}; thus, we find that data from the
$2^\mathrm{nd}$ maximum does not improve the senstivity to discern the
octant of $\theta_{23}$.

The conclusion for three flavor oscillation in this case is the same
as with only one detector: The $2^\mathrm{nd}$ maximum does not help
with the measurement of $\theta_{13}$ or the CP phase, but it enhances
the ability to measure the mass hierarchy. The gains in mass hierarchy
sensitivity could be substantial under favorable conditions, but are
only moderate in practice.

\subsection{Non-standard Interactions}

\begin{figure}[t!]
  \begin{center}
    \includegraphics[width=10cm]{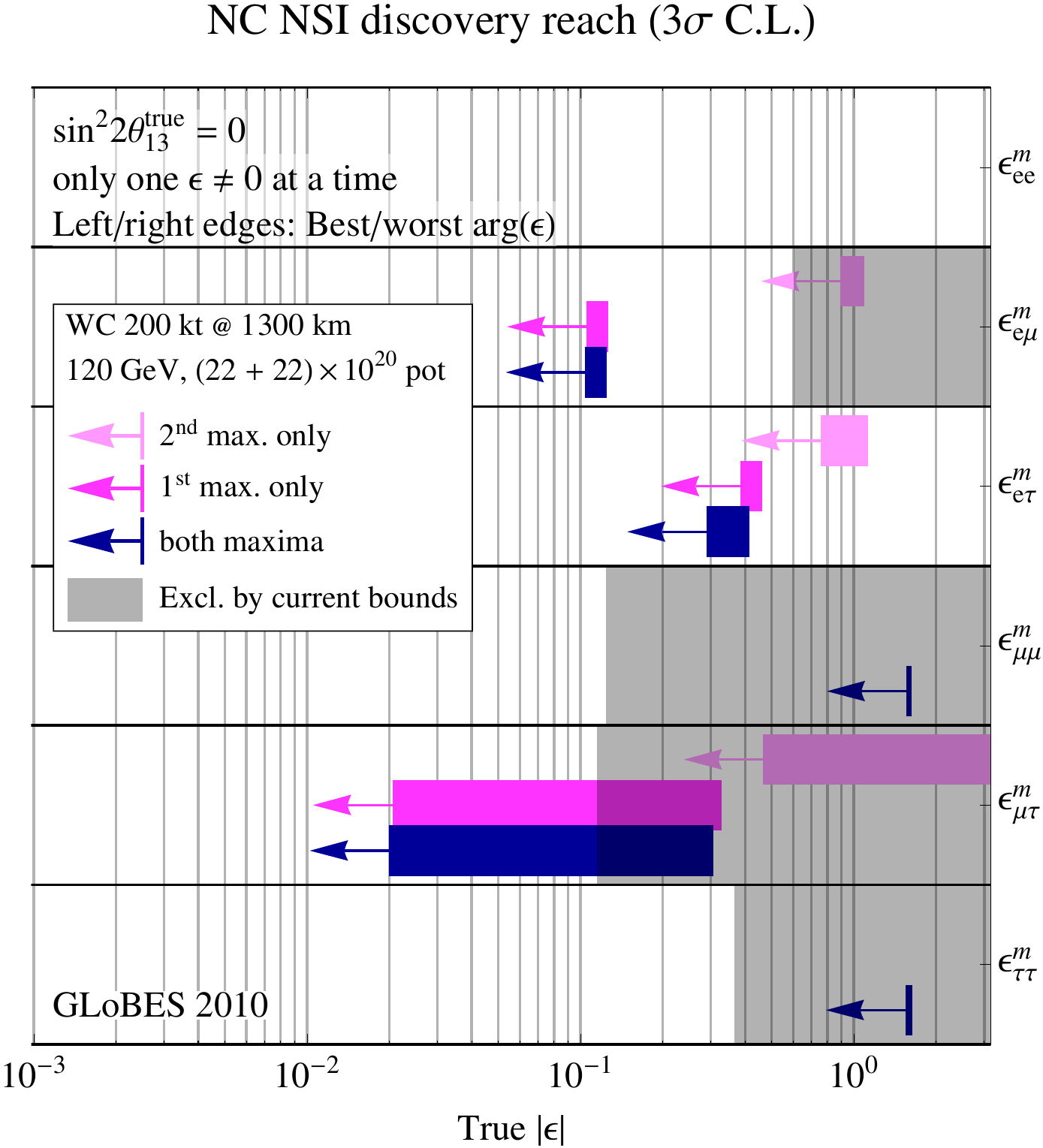}
  \end{center}
  \caption{Discovery reach for neutral-current-like NSI in a WC
    detector (colored bars and arrows), compared to current model-independent
    limits on the different parameters~\cite{Davidson:2003ha,
    GonzalezGarcia:2007ib, Biggio:2009nt} (gray shaded regions).
    The length of the colored bars is due
    to the unknown phase associated with the new interaction.}
  \label{fig:NSI-WC}
\end{figure}

Next we turn our attention to the question whether the $2^\mathrm{nd}$
maximum is useful for new physics searches. For a given set of
oscillation parameters the relative strength of the signals in the
$1^\mathrm{st}$ and $2^\mathrm{nd}$ maximum is well understood within the
standard three flavor oscillation framework, and therefore any deviation should
stem from new physics. In order to be able to perform a quantitative
analysis of this problem, we will restrict the new physics to the form
of non-standard interactions, their underlying physics and
parametrization have been described in section~\ref{sec:nsi}.

In figure~\ref{fig:NSI-WC} the discovery reach for neutral current
like NSI is shown for our standard setup.  For each set of bars, only
one NSI parameter was allowed to be nonzero at a time, {\it i.e.}\ we
do not include correlations between different NSI parameters. The
length of the bars is due to the unknown phase of the non-standard
parameters, whereas the different colors are for different subsets of
the data as explained in the legend. The gray shaded areas indicate the
current model independent bounds on these
parameters~\cite{Davidson:2003ha, GonzalezGarcia:2007ib,
Biggio:2009nt};%
\footnote{We have converted the 90\% C.L. bounds given in
  refs.~\cite{Davidson:2003ha, GonzalezGarcia:2007ib,Biggio:2009nt}
  to the $3\,\sigma$ confidence level assuming Gaussian errors.}
in cases where there is no gray shaded area, the
current bounds are of order one. Note that possible correlations
between $\eps^m_{ee}$ and other parameters are equivalent to
correlations with the matter density, which are included in our
simulations by the matter density uncertainty.  Also, correlations
between $\eps^m_{e\mu}$, $\eps^m_{e\tau}$ and the $\mu$--$\tau$ sector
($\eps^m_{\mu\mu}$, $\eps^m_{\mu\tau}$, $\eps^m_{\tau\tau}$) are small
because $\eps^m_{e\mu}$ and $\eps^m_{e\tau}$ affect mainly the
appearance channel, while $\eps^m_{\mu\mu}$, $\eps^m_{\mu\tau}$ and
$\eps^m_{\tau\tau}$ are most relevant in the disappearance channel,
see {\it e.g.}\ references~\cite{Kopp:2007ne,Kopp:Phd-Thesis}; this has
been shown for the $\eps^m_{e\tau}$--$\eps^m_{\tau\tau}$ correlation
by explicit numerical calculation in Ref.~\cite{Kopp:2008ds}. The
strongest improvement in bounds happens for flavor-changing NSI, but
this improvement is hardly dependent on the data
from the $2^\mathrm{nd}$ oscillation maximum. The results for a liquid
argon detector are very similar and lead to the same conclusion.

Based on our somewhat negative result with respect to the
$2^\mathrm{nd}$ maximum in the case of standard three-flavor
oscillation, we can study the effect of an artificial enhancement of
statistics in the $2^\mathrm{nd}$ maximum also in the presence of
non-standard interactions.  The result of rescaling the flux according
to equation~\ref{eq:param} is shown in the lower right hand panel of
figure~\ref{fig:variation} for the case of NSI in the $e$--$\mu$
sector. The performance optimum occurs at $x_2 < 1$, which implies
that the $2^\mathrm{nd}$ maximum is not useful in this case. The
option of using a second detector to access the $2^\mathrm{nd}$
maximum exists also for NSI studies and the result is shown in the
lower right hand panel of figure~\ref{fig:two-det}.  With a second
detector between $3,000-6,000\,\mathrm{km}$ the sensitivity would be
improved by less than a factor of two. The wide baseline range over which this
improvement happens makes it seem unlikely that this is entirely due
to the $2^\mathrm{nd}$ maximum. Also, in contrast to the case of the
mass hierarchy measurement, where there was an improvement both for
rescaling the flux and considering a second detector, here we see
improvement only for the second detector, which points to overall
increased matter effects, standard and non-standard, as the source of
this improvement. Qualitatively a similar improvement is obtained by
just using one detector at a longer baseline as shown in
figure~\ref{fig:baseline}. In any case, the improvement is relatively
moderate.

The previous statement about the relative unimportance\footnote{ While
  there may be correlations between the various parameters, the fact
  that all relevant $\epsilon$ involving $\mu$-type flavor are tightly
  constrained, should make the correlations practically negligible.}
of correlations between NSI parameters does not hold if one allows for
the simultaneous presence of non-standard effects in the neutrino production, propagation,
and detection processes. In particular, in this case the so called confusion theorem
obtains~\cite{Huber:2002bi}: Assume that $\theta_{13} = 0$, but there are
charged current NSI between $\nu_\tau$ and electrons in the
detector ($\epsilon_{\tau e}^d \neq 0$), and neutral current NSI
between $\nu_e$ and $\nu_\tau$ in the propagation ($\epsilon_{e\tau}^m
\neq 0$).\footnote{The original confusion theorem was derived in
  the context of a neutrino factory, where the appearance signal stems
  from $\nu_e\rightarrow\nu_\mu$ oscillations. There, it is nonzero $\eps^m_{e\tau}$
  together with a CC-like NSI in the \emph{source} ($\epsilon^s_{e\tau} \neq 0$)
  which causes the confusion. Here, we are considering the $T$-conjugate
  oscillation channel, and hence we need a CC-like NSI in the detection
  process instead.} If we furthermore assume that the parameters obey
the relation
\begin{equation}
  \epsilon_{\tau e}^d=r\epsilon_{e\tau}^m \,,
  \label{eq:confusion-1}
\end{equation}
with $r$ being an order one parameter determined by whether the NSI
couple to quarks or leptons, then the event rate spectra for both
neutrinos and anti-neutrinos in the $\nu_\mu\rightarrow\nu_e$ channel
are the same as for standard three-flavor oscillations with
\begin{equation}
  \sin^2\theta_{13}=r^2 (\epsilon_{e\tau}^m)^2\frac{1+\cos2\theta_{23}}{2}\,.
  \label{eq:confusion-2}
\end{equation}
This is the confusion theorem. Subsequently, it was discovered that
for sufficiently high beam energies, muons from the decay of $\tau$
from $\nu_\tau$ charged current interactions can be used to resolve
the confusion at least for parts of the parameters
space~\cite{Campanelli:2002cc}. For the setup considered here, the
average beam energy is close to $m_\tau$, and therefore
$\tau$-production in charged current interactions from $\nu_\tau$ will
be strongly suppressed, and therefore no muons from $\tau$-decays will
be observed. Thus, the confusion theorem should apply. Still,
it is important to note that equations~\ref{eq:confusion-1} and \ref{eq:confusion-2}
were obtained from a perturbative expansion of the oscillation probability. This expansion
is strictly valid only for energies around the $1^\mathrm{st}$ maximum.
Thus the question arises to which degree the confusion theorem applies
to the wide band scenario considered here.

\begin{figure}[t!]
  \begin{center}
    \includegraphics[width=\textwidth]{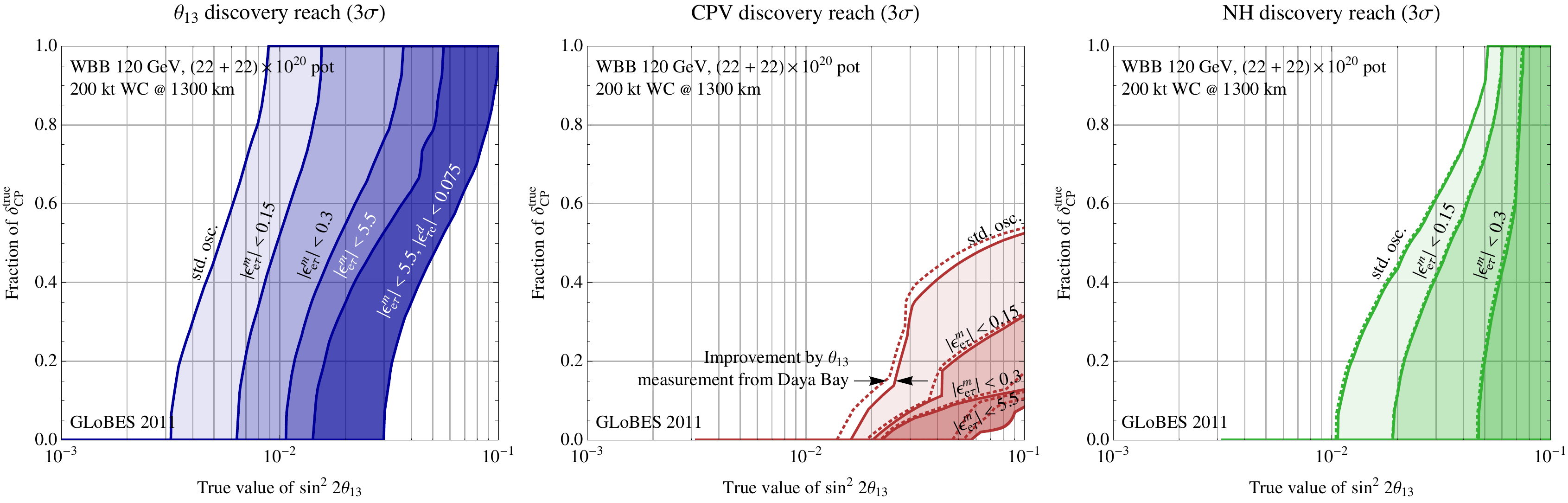}
  \end{center}
  \caption{Discovery reach for $\sin^22\theta_{13}$, $\deltaCP$ and
    the mass hierarchy in the presence of NSI in propagation and
    detection, specifically $\eps^m_{e\tau}$ and $\eps^d_{\tau e}$.
    In  all panels, the leftmost line is the usual three flavor
    oscillation result and the other lines are obtained by successively
    allowing NSI up to the limit indicated by the labels next to each
    line. The rightmost curves, corresponding to the current $3\,\sigma$
    limits on the NSI parameters~\cite{Davidson:2003ha,Biggio:2009nt},
    lie outside the plot for $\deltaCP$ and the mass hierarchy.
    The dashed lines in the middle and right panels are the
    corresponding results if one assumes $\theta_{13}$ to be
    constrained by a measurement at Daya Bay.
    \label{fig:confusion}}
\end{figure}

A partial answer is shown in figure~\ref{fig:confusion}, where the
sensitivity to $\sin^22\theta_{13}$, CP violation and the mass
hierarchy is depicted for various levels of NSI.  The different lines
are obtained by allowing successively larger values of
$|\epsilon^m_{e\tau}|$ and $|\epsilon^d_{\tau e}|$ in the fit, as indicated in the
plots.  The largest values used in figure~\ref{fig:confusion} correspond
to the current $3\,\sigma$ bounds according to
references~\cite{Davidson:2003ha, Biggio:2009nt}.%
\footnote{Again, we have converted 90\% C.L.\ limits to $3\,\sigma$
  constraints assuming Gaussian errors.}
In the left panel ($\theta_{13}$ discovery reach) and in the right panel
(discovery reach for the normal mass hierarchy), we allow $\epsilon^m_{e\tau}$
and $\epsilon^d_{\tau e}$ to be complex with arbitrary phases, while in
the middle panel (discovery reach for CP violation), we assume the phases
of the NSI parameters to be $0$ or $\pi$ since we want to consider only CP
conserving solutions in the fit. The rightmost line in the
left hand panel of figure~\ref{fig:confusion} confirms the validity of
the confusion theorem in equations~\ref{eq:confusion-1} and \ref{eq:confusion-2}: We
indeed observe a deterioration of the sensitivity by nearly an order of magnitude in
$\sin^22\theta_{13}^{\rm true}$. However, at the same time we see that
$\epsilon^m_{e\tau}$ alone accounts for most of this deterioration
since the difference between the rightmost line and the line next to
it is relatively small. Thus, the confusion theorem seems to apply in
essence, but in practice the small number of events around the
sensitivity limit does not require the presence of NSI both in
propagation and detection, because spectral information is not
statistically significant. For the same reason, the information from
the $2^\mathrm{nd}$ maximum plays no role, since at the sensitivity limit
the event sample from the $2^\mathrm{nd}$ maximum is statistically not
significant. We have performed
the same scaling analysis as presented in figure~\ref{fig:variation}
also in this case and find that the $2^\mathrm{nd}$ maximum is not
useful in controlling the effects of NSI in propagation and detection.

In the middle panel of figure~\ref{fig:confusion} we show the impact
of NSI on the ability to discover CP violation, and the result exhibits
the same qualitative features as the one for the discovery of
$\theta_{13}$. At a quantitative level, this measurement is more
sensitive to the deleterious effects of NSI since it relies on
smaller, more difficult signatures also in the standard oscillation
case.  Therefore, we observe a complete loss of discovery potential
for values of the NSI an order of magnitude below the current bounds.
One may speculate that using a precision measurement of $\theta_{13}$
by the Daya Bay experiment could mitigate the correlation with NSI,
however, as the dashed lines conclusively demonstrate, this is note the
case. Thus, at the current level of analysis we are forced to conclude
that the ability to discover leptonic CP violation in the next
generation of superbeam experiments is \emph{not} robust with respect
to the presence of new physics. Neither a $\theta_{13}$ measurement by
Daya Bay nor events from the $2^\mathrm{nd}$ maximum can resolve this
problem. A direct measurement of $\epsilon^m_{e\tau}$ in a short-baseline
neutral current neutrino scattering experiment to improve
the upper limits is out of question, since this would require to
measure the flavor of the outgoing neutrino. As shown in figure~4 of
reference~\cite{Huber:2002bi}, combining different baselines does not
affect the validity of the confusion theorem and therefore does not
present a viable strategy to address the confusion problem either.

In the rightmost panel of figure~\ref{fig:confusion} we show how much
the discovery reach for the mass hierarchy is diminished, and the
result is equally drastic as in the case of CP violation.

A near detector can help to improve bounds on NSI in the neutrino
production and detection processes, but our results show that NSI in
propagation alone are sufficient to cause serious problems for the
measurement of the standard oscillation parameters. Besides, a direct
measurement of $\epsilon^d_{\tau e}$ is a difficult proposition for a
number of reasons. First, there are no $\nu_\tau$ in the beam, so the
only way of constraining $\epsilon^d_{\tau e}$ would be to constrain
$\epsilon^s_{e\tau}$ instead and to make use of the fact that the
two parameters are usually related since they can both arise from
the same non-standard coupling between two light quarks, an electron,
and a $\nu_\tau$~\cite{Kopp:2007ne}. However, this relation between
$\epsilon^d_{\tau e}$ and $\epsilon^s_{e\tau}$ is not model-independent.
For instance, a parity-conserving non-standard operator can lead to
nonzero $\epsilon^d_{\tau e}$, but will not contribute to neutrino
production in pion decay, so that $\epsilon^s_{e\tau}
= 0$~\cite{Kopp:2010qt}. Moreover, even measuring
$\epsilon^s_{e\tau}$ is very challenging because the initial flux of
electron neutrinos in an LBNE-like beam is very small, less
than 1\%, the kinematic suppression of $\tau$-production is large
with the available beam energy, and $\tau$-identification is
notoriously difficult and typically has a low efficiency. For these
reasons, we conclude that near detectors will not solve the problem
of possible confusion between standard oscillations and NSI. The specific
setup considered here is in some sense a best case scenario, since
it has relatively high statistics, a lot of spectral information, and
makes use of two oscillation maxima.  There is no reason to
expect that experiments like T2K and NO$\nu$A will be less affected
by the confusion problem, quite the contrary, as shown in
reference~\cite{Kopp:2007ne}.

\section{Summary and conclusions}
\label{sec:summary}

The goal of this paper is to  quantitatively understand the benefits,
or lack thereof, of studying two oscillation maxima simultaneously
in a long-baseline neutrino oscillation experiment. To this end we have
chosen a specific example of experimental setup, which closely
resembles the current plans for the Fermilab--DUSEL Long Baseline
Neutrino Experiment (LBNE). There are two ways to access the
$2^\mathrm{nd}$ oscillation maximum: either using a broad neutrino energy
spectrum to cover the $1^\mathrm{st}$ and $2^\mathrm{nd}$ maximum in
the same detector, or using two detectors at different baselines in
the same beam. In LBNE, the natural method is to use the same detector
and a wide energy spectrum. For this approach we found that there is no
apparent benefit from using the $2^\mathrm{nd}$ maximum
(figure~\ref{fig:variation}). This remains true even if it were
possible to shift a larger portion of the total neutrino flux into
the energy range of the $2^\mathrm{nd}$ oscillation maximum.
The measurement of the mass hierarchy does
improve with events from the $2^\mathrm{nd}$ maximum, but the
improvement is limited, and since it would come at the cost of
losing events in the $1^\mathrm{st}$ maximum, which in turn negatively
impacts the other measurements, a trade-off between these conflicting
requirements has to be found. We have also investigated the option of using
the same beam but two detectors at different baselines, which is the
natural option for extensions of
T2K~\cite{Ishitsuka:2005qi,Dufour:2010vr,Badertscher:2008bp}. The
results are similar to the previous case for the measurement of
$\theta_{13}$ and CP violation, while the improvement in the sensitivity
to the mass hierarchy is somewhat stronger in this case due to larger
matter effects at the longer baseline (figure~\ref{fig:two-det}).
As far as the possible detection of new physics---parametrized here in the
framework of neutral current non-standard interactions (NSI)---is concerned, the
sensitivity improves slightly for a second detector at a longer baseline.
Note that neutral current NSI measurements prefer longer baselines in
general since they essentially correspond to new neutrino
matter effects, and matter effects are larger at long baseline. Therefore
a similar improvement can also be observed for a single detector setup with a
longer baseline (figure~\ref{fig:baseline}).

Finally, we have revisited the so called confusion theorem, which has been
discovered in the context of neutrino factories~\cite{Huber:2002bi}.
The confusion theorem states that certain combinations of charged and
neutral current NSI modify the neutrino oscillation probability in the same
way as a nonzero value of $\theta_{13}$ does. In particular, the effects of
the NSI parameter $\epsilon^m_{e\tau}$ are problematic since this parameter is only weakly
constrained by an $\mathcal{O}(1)$ bound.  In other words, in the absence of
sufficiently strong bounds on NSI, it is very hard to establish a
bound on $\theta_{13}$. In the context of neutrino factories, muons
from $\tau$ decays have proven to be a loophole in the confusion
theorem~\cite{Campanelli:2002cc}, but in a superbeam experiment the beam
energy is so low that $\tau$-production is kinematically suppressed
and the arguments from reference~\cite{Campanelli:2002cc} do not apply,
as has been shown in the context of T2K and NO$\nu$A~\cite{Kopp:2007ne}.
In the present work we have extended those earlier results to
superbeam experiments which have events from the $1^\mathrm{st}$ and
$2^\mathrm{nd}$ oscillation maximum. We have found that the
confusion theorem holds (figure~\ref{fig:confusion}) and the
sensitivity to $\sin^22\theta_{13}$ deteriorates by one order of
magnitude if the possibility of NSI is taken into account. Data from
the $2^\mathrm{nd}$ maximum has no effect on this conclusion since it
is statistically not significant enough. Moreover, we have shown that
even the effect of neutral current NSI ($\epsilon_{e\tau}^m \neq 0$)
alone is sufficient to seriously impact the sensitivity to
$\sin^22\theta_{13}$. This is an especially problematic limitation
since $\epsilon_{e\tau}^m$ cannot be constrained by a near detector
measurement. For CP violation measurements, the possibility of complex
$\epsilon_{e\tau}^m$ has been considered and the impact is dramatic:
Even for $\epsilon_{e\tau}^m$ an order of magnitude below the current
bound a complete loss of sensitivity ensues. Reactor experiments
will not be affected by $\epsilon_{e\tau}^m$ and thus
can provide a clean measurement of $\sin^22\theta_{13}$, but we
have shown that even precise knowledge of $\theta_{13}$ from a reactor
experiment does not solve the problem. The presence of data from the
$2^\mathrm{nd}$ maximum does not provide immunity from the confusion
theorem and it has an overall very small quantitative impact. This
remains true even if the majority of the neutrino flux were shifted into
the $2^\mathrm{nd}$ maximum. In comparison, T2K and NO$\nu$A have
smaller statistics and observe a much smaller range in $L/E$.
Therefore, the impact of the confusion theorem is more severe in these
experiments~\cite{Kopp:2007ne}.

To put our conclusion on the possible impact of large NSI into
perspective, we should emphasize that NSI large enough to be
problematic for T2K, NO$\nu$A, or LBNE, are \emph{not} a generic
prediction of extensions of the Standard Model. In particular, from a
model-builder's point of view, new particles at or above the
electroweak scale are rather unlikely to have a sizable effect on the
neutrino sector~\cite{Antusch:2008tz,Gavela:2008ra}. On the other
hand, new physics at a low scale may still lead to large
NSI~\cite{Joshipura:2003jh,Nelson:2007yq}, and since neutrino physics
has taught us in the past that theorists' prejudices may be wrong, one
cannot discard such possibilities.

In summary, we find that only the determination of the mass hierarchy benefits
slightly from using the $2^\mathrm{nd}$ oscillation maximum in a long baseline
neutrino oscillation experiment.  All other measurements, including
non-standard neutrino interactions, are not improved compared to the case where
only events from the $1^\mathrm{st}$ maximum are used. We confirm that the
``confusion theorem'', which states that certain types of NSI can mimic the
effects of nonzero $\theta_{13}$, remains valid even if data from the
$2^\mathrm{nd}$ maximum is available. In particular, we have shown that the
presence of a non-standard coupling between electron neutrinos and $\tau$
neutrinos with a complex coefficient $\epsilon_{e\tau}^m$ can completely
destroy the sensitivity to CP violation and the mass hierarchy.  Since it is
known from the literature~\cite{Huber:2002bi} that combining different
baselines does not alleviate this problem, and we have explicitly shown that
adding reactor neutrino data does not work to this end either, it seems that
there are no simple remedies for the confusion problem.

\subsubsection*{Acknowledgments}
We are indebted to the members of the LBNE collaboration,
especially Mary Bishai, Bonnie Fleming, Roxanne Guenette, Gina
Rameika, Lisa Whitehead, and Geralyn `Sam' Zeller for providing
invaluable information on the parameters of the planned Fermilab
neutrino beams, DUSEL detectors, and other aspects of the LBNE experiment.
We are also grateful to Mattias Blennow for some very useful discussions,
especially during the early stages of this project. PH would like to
acknowledge the warm hospitality at the {\it Astroparticle Physics - A
Pathfinder to New Physics} workshop at the KTH in Stockholm, during
which this project was conceived, and the $\nu$TheME institute at CERN,
where it was brought to completion. 

Fermilab is operated by Fermi Research Alliance, LLC under Contract
No.~\protect{DE-AC02-07CH11359} with the US~Department of Energy. This
work has been in part supported by the US~Department of Energy under
award number \protect{DE-SC0003915}.

\begin{appendix}

\section{Possible alternative setups}
\label{app:alt}

\begin{figure}[t]
  \begin{center} \includegraphics[width=\textwidth]{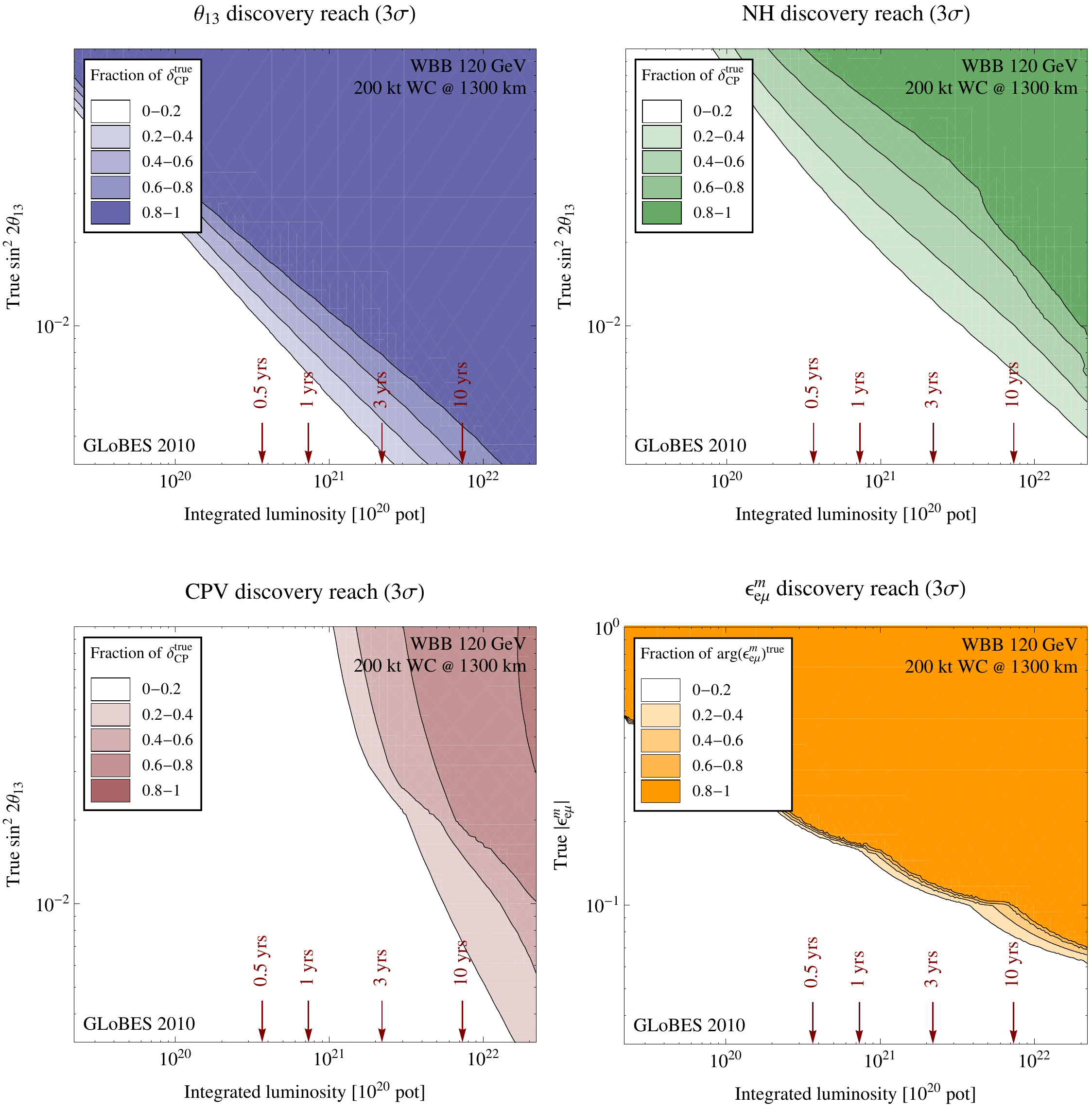}
  \end{center}
  \caption{WBB discovery reach as a function of the exposure. The
    vertical, red arrows indicate the running time required with the
    standard value of $7.3\times
    10^{20}\,\mathrm{pot}\,\mathrm{y}^{-1}$ at $120\,\mathrm{GeV}$.
    All plots are for a $200\,\mathrm{kt}$ WC detector.  The curves
    indicate different CP fractions as given in the legend.}
  \label{fig:exposure}
\end{figure}

In this appendix we study simple variations of the basic experimental
setup considered in this paper. The
results given here will allow to extrapolate the effects of changes in
the exposure and baseline. In figure~\ref{fig:exposure} we show the
discovery reach for standard oscillation as well as NSI as a function
of exposure. Obviously, the higher the exposure the better the
sensitivity. Discovery of CP violation has the largest demand for high
exposure and conversely is most at risk if the luminosity were to turn
out smaller than expected.

\begin{figure}[t]
  \begin{center}
\includegraphics[width=\textwidth]{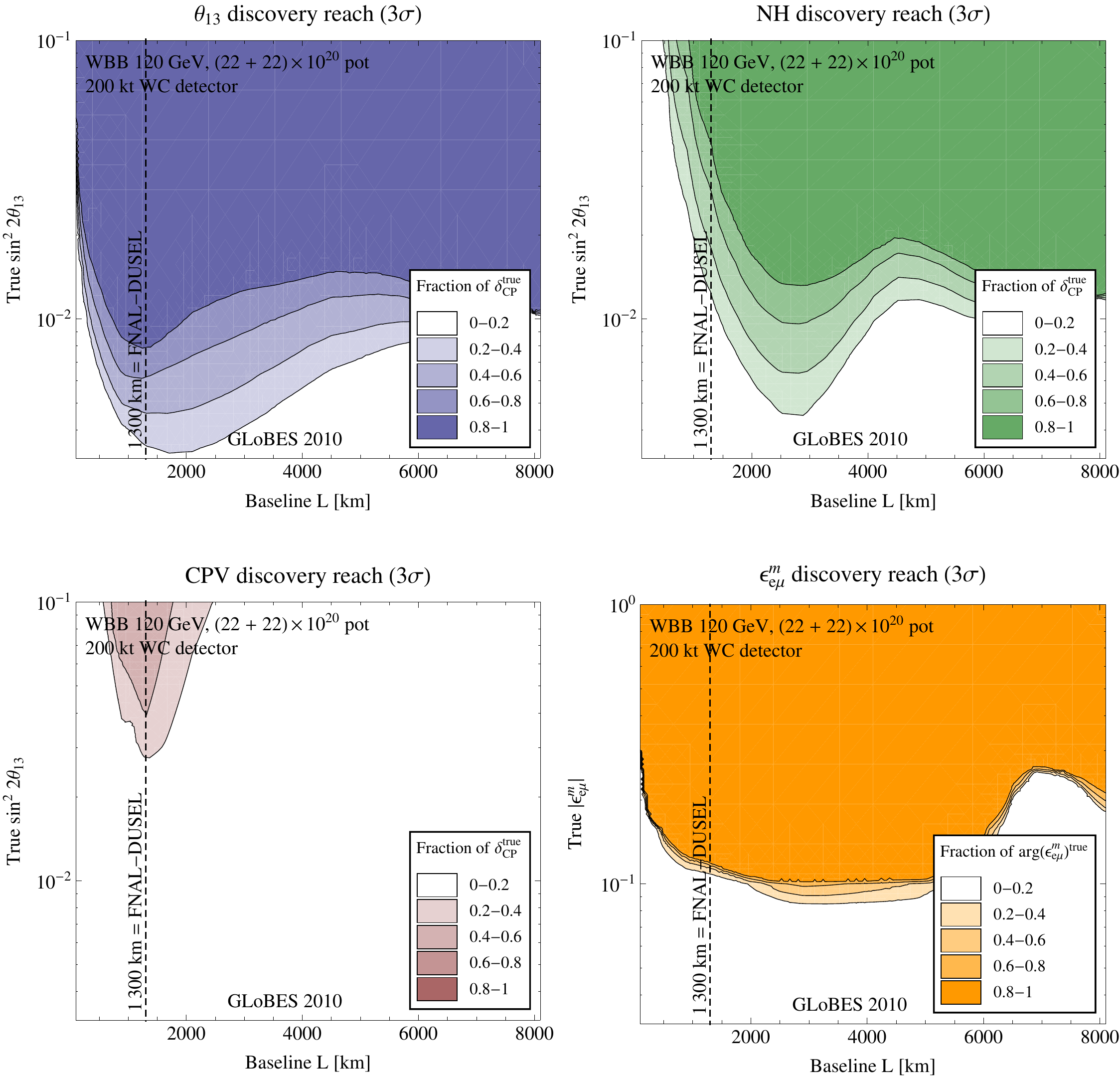}
  \end{center}
  \caption{\label{fig:baseline} WBB discovery reach as a function of
    the baseline using a 200~kt WC detector. The lines and the
    corresponding shades are iso-contours of CP fraction with values
    given in the legend. The four panels are in the top row:
    $\theta_{13}$ discovery reach (left) and mass hierarchy discovery
    reach (right). In the lower row we have: discovery reach for CP
    violation (left) and discovery reach for the NSI parameter
    $\epsilon_{e\mu}^m$ as one example of NSI sensitivities. All
    results are shown at the $3\,\sigma$ confidence level.}
  \label{fig:Lth13FracDelta}
\end{figure}

Figure~\ref{fig:baseline} illustrates the dependence of the discovery
reaches for $\theta_{13}$, mass hierarchy, CPV discovery and the NSI
parameter $\epsilon_{e\mu}^m$ as a function of the baseline.
The lines and corresponding shades are iso-contours of CP fraction.
The optimum occurs for CP violation and the discovery of $\theta_{13}$
around $1,500\,\mathrm{km}$, whereas the optimum for the discovery of
the mass hierarchy and $\epsilon_{e\mu}^m$ is around
$2,200\,\mathrm{km}$. This result confirms that $L=1,300\,\mathrm{km}$
is a reasonable comprise for the neutrino beam assumed in this paper,
and slightly longer baselines around $1,600\,\mathrm{km}$ would perform
very similarly.

\end{appendix}


\bibliographystyle{apsrev4-1}
\bibliography{references}

\end{document}